\documentclass[12pt,preprint]{aastex}
\usepackage{graphicx}
\usepackage{natbib}
\usepackage{url}

\newcommand{\zabs}{\ensuremath{z_\textrm{\scriptsize abs}}}
\newcommand{\zem}{\ensuremath{z_\textrm{\scriptsize{em}}}}
\newcommand{\dmm}{\Delta\mu/\mu}
\newcommand{\kms}{km\,s$^{-1}$}
\newcommand{\ms}{m\,s$^{-1}$}
\newcommand{\nuprime}{\nu\,'}
\newcommand{\bx}{\mbox{B-X$(0-0)$}}
\newcommand{\cx}{\mbox{C-X$(0-0)$}}
\newcommand{\ex}{\mbox{E-X$(0-0)$}}
\newcommand{\jj}{J'J''}
\newcommand{\vv}{\nu' \nu''}

\newcommand{\apjss}{Astroph.\ J.\ Supp.\ Ser.}

\newcommand{\cp}{Chem.\ Phys.}
\newcommand{\jms}{J. Mol.\ Spectrosc.}

\newcommand{\eujpd}{Eur.\ J. \ Phys. \ D}
\newcommand{\revmodphys}{Rev. Mod. Phys.}

\newcommand{\lrr}{Living Rev. Relat.}

\newcommand{\soviet}{Sov. J. Exp. Theor. Phys.}

\newcommand{\molphys}{Mol. Phys.}

\newcommand{\AcPSn}{AcPSn}

\shorttitle{A $\mu$-variation constraint from CO absorption}
\shortauthors{Dapr\`a M. et al.}

\begin{document}

\title{Constraint on a cosmological variation in the proton-to-electron mass ratio from electronic CO absorption}

\author{M. Dapr\`a}
\affil{Department of Physics and Astronomy, LaserLaB, Vrije Universiteit University, De Boelelaan 1081, 1081 HV Amsterdam, The Netherlands}
\email{m.dapra@vu.nl}

\and

\author{M. L. Niu}
\affil{Department of Physics and Astronomy, LaserLaB, Vrije Universiteit, De Boelelaan 1081, 1081 HV Amsterdam, The Netherlands}

\and

\author{E. J. Salumbides\altaffilmark{1}}
\affil{Department of Physics and Astronomy, LaserLaB, Vrije Universiteit, De Boelelaan 1081, 1081 HV Amsterdam, The Netherlands}

\and

\author{M. T. Murphy}
\affil{Centre for Astrophysics and Supercomputing, Swinburne University of Technology, Melbourne, Victoria 3122, Australia}

\and

\author{W. Ubachs}
\affil{Department of Physics and Astronomy, LaserLaB, Vrije Universiteit, De Boelelaan 1081, 1081 HV Amsterdam, The Netherlands}
\email{w.m.g.ubachs@vu.nl}

\altaffiltext{1}{Department of Physics, University of San Carlos, Cebu City 6000, Philippines}

\begin{abstract}
Carbon monoxide (CO) absorption in the sub-damped Lyman-$\alpha$ absorber at redshift \mbox{$\zabs \simeq 2.69$}, toward the background quasar SDSS J123714.60+064759.5 (J1237+0647), was investigated for the first time in order to search for a possible variation of the proton-to-electron mass ratio, $\mu$, over a cosmological time-scale. The observations were performed with the Very Large Telescope/Ultraviolet and Visual Echelle Spectrograph with a signal-to-noise ratio of 40 per 2.5 \kms\ per pixel at \mbox{$\sim 5000$ \AA}. Thirteen CO vibrational bands in this absorber are detected: the A$^{1}\Pi$ - X$^{1}\Sigma^{+}$ ($\nuprime$,0) for $\nuprime = 0 - 8$, B$^{1}\Sigma^{+}$ - X$^{1}\Sigma^{+}$ (0,0), C$^{1}\Sigma^{+}$ - X$^{1}\Sigma^{+}$ (0,0), and E$^{1}\Pi$ - X$^{1}\Sigma^{+}$ (0,0) singlet-singlet bands and the d$^{3}\Delta$ - X$^{1}\Sigma^{+}$ (5,0) singlet-triplet band. An updated database including the most precise molecular inputs needed for a $\mu$-variation analysis is presented for rotational levels \mbox{$J = 0 - 5$}, consisting of transition wavelengths, oscillator strengths, natural lifetime damping parameters, and sensitivity coefficients to a variation of the proton-to-electron mass ratio. A comprehensive fitting method was used to fit all the CO bands at once and an independent constraint of \mbox{$\dmm = (0.7 \pm 1.6_{stat} \pm 0.5_{syst}) \times 10^{-5}$} was derived from CO only. A combined analysis using both molecular hydrogen and CO in the same J1237+0647 absorber returned a final constraint on the relative variation of \mbox{$\dmm = (-5.6 \pm 5.6_{stat} \pm 3.1_{syst}) \times 10^{-6}$}, which is consistent with no variation over a look-back time \mbox{of $\sim 11.4$ Gyrs}.
\end{abstract}


\section{Introduction} \label{sec:intro}
The Standard Model of particle physics depends on a number of parameters that cannot be explained from the model itself. These parameters, including for example the fine-structure constant \mbox{$\alpha = e^{2} / (4 \pi \epsilon_{0} \hbar c)$}, and the proton-to-electron mass ratio \mbox{$\mu \equiv M_{P}/m_{e}$}, are referred to as fundamental constants and are assumed to be spacetime invariant. Whether they are really constant or whether they undergo variations over time is a question that became subject of observation when \cite{Savedoff1956} established the alkali-doublet method to compare galaxy values of physical constants with local values. Subsequently, \cite{Thompson1975} suggested that a cosmological variation of $\mu$ could be probed using molecular hydrogen (H$_{2}$) absorption in quasar spectra. A number of theories predicting a variation of constants have been proposed \citep[an extensive review has been given by][]{Uzan2011}, often associated with forces beyond the four known in the Standard Model or with extra dimensions beyond the $3+1$ presently assumed. It is noted that a variation of the fundamental constants implies a violation of Einstein's equivalence principle, which is a basic assumption of General Relativity.

A sensitive search for a cosmological variation of dimensionless fundamental constants $\alpha$ and $\mu$ is possible via the measurement of atomic and molecular absorption lines detected at high redshifts in the line-of-sight towards quasars, using high resolution spectroscopic observations in the optical and in the radio domain. The fine-structure constant $\alpha$ was investigated by recording spectroscopic lines of atoms \citep{Dzuba1999}, looking for variations in the temporal \citep{Webb1999} and spatial domains \citep{Webb2011}. The cosmological variation of proton-to-electron mass ratio $\mu$, which is sensitive to the ratio of the chromodynamic to the electroweak scale \citep{Flambaum2004}, can be probed using molecular absorption \citep{Jansen2011b,Jansen2011,Jansen2014}. Various searches for $\mu$-variation via observation of ammonia \citep[NH$_{3}$,][]{Flambaum2007,Murphy2008,Kanekar2011} and methanol \citep[CH$_{3}$OH,][]{Bagdonaite2013a,Bagdonaite2013b,Kanekar2015} spectral lines at intermediate redshifts \mbox{$z < 1$}, yielded a constraint on $\dmm$ on the order of \mbox{$\sim 10^{-7}$} at the $1\sigma$ level. 

Molecular hydrogen, the most abundant molecule in the Universe, is used to investigate $\mu$-variation in absorbing systems  at redshifts \mbox{$\zabs > 2$}, for which the Lyman and Werner bands fall into the optical band. The analysis of H$_{2}$ absorption was performed in nine systems detected in the range $z = 2.05 - 4.22$ \citep{Reinhold2006,King2008,Thompson2009,Malec2010,King2011,Weerdenburg2011,Wendt2011,Bagdonaite2012,Wendt2012,Rahmani2013,Bagdonaite2014,Vasquez2014,Bagdonaite2015,Dapra2015}, delivering an averaged constraint (at the 3$\sigma$ level) of $|\dmm|$ of \mbox{$< 5 \times 10^{-6}$} \citep{Ubachs2016}. The analysis procedure used in the H$_{2}$ method relies on the calculation of sensitivity coefficients for the hydrogen molecule, i.e.\ how much each transition shifts in wavelength for a given change in $\mu$ \citep{Varshalovich1993,Ubachs2007}. In addition, some molecules exhibit sensitivities to both fundamental constants $\alpha$ and $\mu$. \cite{Tzanavaris2005} combined \ion{H}{1} 21-cm lines with UV metal absorption lines to estimate the time variation of the combination $\alpha^2 g_p \mu$, with $g_p$ the dimensionless proton $g$-factor. \cite{Kanekar2012} observed OH microwave lines and searched for a combined variation of both constants $\alpha$ and $\mu$.

Carbon monoxide (CO) is the second most abundant molecule in the Universe and it is one of the best studied molecules in spectroscopy. Being sensitive to $\mu$-variation, \cite{Salumbides2012} proposed to use its electronic \mbox{A$^{1}\Pi$ - X$^{1} \Sigma^{+}$} system as a suitable target to constrain a cosmological variation of $\mu$. So far, optical absorption bands of CO are detected in six absorption systems at redshifts $\zabs > 1$: SDSS J160457.50+220300.5 \citep{Noterdaeme2009}, SDSS J085726+185524, SDSS J104705.75+205734.5, SDSS J170542+354340 \citep{Noterdaeme2011}, SDSS J143912.04+111740.5 \citep{Srianand2008,Noterdaeme2008}, and SDSS J123714.60+064759.5 \citep{Noterdaeme2010}. The absorbing system at $\zabs \simeq 2.69$ towards quasar SDSS J123714.60+064759.5, hereafter J1237+0647, is an exemplary absorbing system that contains high quality spectra of both H$_{2}$ and CO, providing a case for a combined analysis of $\mu$-variation using the two molecules. 

In the present study, the electronic CO absorption in this absorber is investigated in order to obtain a constraint on a temporal $\mu$-variation over cosmological timescales. The observations used in this work are listed in {Section \ref{sec:observations}}, while the molecular database containing the parameters used to build the absorption model of CO is presented in \mbox{Section \ref{sec:co}}. The model, as well as the comparison with a previous constraint derived from the analysis of H$_{2}$ absorption in the same system \citep{Dapra2015} is presented  in  \mbox{Section \ref{sec:model}}, and the analysis of the systematic uncertainty is given in \mbox{Section \ref{sec:systematics}}. The results are summarized in \mbox{Section \ref{sec:conclusion}}.

\section{Observations} \label{sec:observations}
The dataset used in this work is gathered from four different observing programs carried out between 2009 and 2014 using the Ultraviolet and Visual Echelle Spectrograph (UVES) mounted on the 8.2m Very Large Telescope (VLT) at Paranal, Chile \citep{Dekker2000}. Three programs were performed in service mode; two in March-April 2009 (082.A-0544(A) and 083.A-0454(A), PI Ledoux), of which an analysis was reported by \cite{Noterdaeme2010} and which were retrieved from the ESO archive\footnote{\url{http://archive.eso.org/eso/eso_archive_main.html}} for the present re-analysis, and one between March and June 2014 (093.A-0373(A), PI Ubachs). The program 091.A-0124(A), PI Ubachs, was run in visitor mode in May 2013. 

Exposures taken in 2009 have only the standard ThAr calibration taken at the end of the night, while exposures taken under programs 091.A-0124(A) and 093.A-0373(A) were followed by an attached ThAr calibration exposure and a `supercalibration' exposure of an asteroid or a solar twin taken immediately after the quasar exposure, without allowing for any change in the instrument parameters \citep{Whitmore2015,Dapra2015}. The raw data were bias corrected, flat fielded and their flux was extracted using the Common Pipeline Language version of the UVES pipeline. The wavelength calibration was performed using ThAr lamp exposures. The custom software \textsc{UVES\_popler} \citep{UVESpopler} was used after the standard reduction procedure to combine the echelle orders into the 1D final spectrum and to remove bad pixels and spectral artifacts as well as to fit the continuum using low-order polynomials \citep{Bagdonaite2014,Dapra2015}. The four observational programs return a total of 20 hrs of integration on the target, with a slit width of 1.0 arcsec, a typical seeing of \mbox{$\sim 0.9$ arcsec}, a binning of \mbox{$2 \times 2$}, yielding a resolving power of \mbox{$\lambda/\Delta\lambda \sim 40000$}. The spectrum of J2137+0647 covers the wavelengths from 3290 to \mbox{9600 \AA}, with a signal-to-noise ratio (S/N) of $\sim 13$ per \mbox{2.5 \kms\ per pixel at $\sim 4000$ \AA}, in the blue arm of UVES, and $\sim 40$ per \mbox{2.5 \kms\ per pixel at $\sim 5000$ \AA}, in its red arm.

Quasar exposures taken between 2013 and 2014 were recorded with a supercalibration exposure and were processed following the supercalibration method used by \cite{Dapra2015} in order to correct for wavelength calibration distortions (see \mbox{Section \ref{sec:systematics}}). Some Ceres exposures were taken in 2009 (program 080.C-0881(B), PI Dumas), in a time separation of one week maximum from the quasar observations using only the blue arm of UVES. These exposures were used to supercalibrate the J1237+0647 spectrum at wavelengths shorter than the \mbox{Lyman-$\alpha$} emission feature of the quasar.

\section{CO molecular data} \label{sec:co}
In this section the existing molecular data of the electronic absorption systems of carbon monoxide relevant for quasar absorption studies are reviewed and collected. An extensive data compilation of the CO electronic transitions had been published by \cite{Morton1994} some two decades ago, but in the mean time improved spectroscopic data have been produced, in particular through the accurate wavelength calibrations by laser-based methods and VUV synchrotron absorption studies \citep{Salumbides2012,Niu2013,Niu2015,Niu2016}. Based on these studies, an updated perturbation analysis was performed for excited states of singlet and triplet character. These perturbations also cause an intensity borrowing phenomenon and affect the rotational line strengths. For this reason also some aspects of rotational line strengths, as discussed by \cite{Larsson1983} and \cite{Morton1994}, are re-evaluated for the calculation of line strengths of perturbed lines.

\cite{Noterdaeme2010} reported CO absorption from ten bands in the absorbing system towards J1237+0647. The detected bands belong to three different systems: the \mbox{A$^{1}\Pi$ - X$^{1}\Sigma^{+}$ ($\nuprime$,0)} for $\nuprime=0-7$, \mbox{C$^{1}\Sigma^{+}$ - X$^{1}\Sigma^{+}$ (0,0)}, and \mbox{d$^{3}\Delta$ - X$^{1}\Sigma^{+}$ (5,0)} bands. In the present re-analysis of J1237+0647 data, additional absorption features associated with the \mbox{A$^{1}\Pi$ - X$^{1}\Sigma^{+}$ (8,0)}, \mbox{B$^{1}\Sigma^{+}$ - X$^{1}\Sigma^{+}$ (0,0)}, and \mbox{E$^{1}\Pi$ - X$^{1}\Sigma^{+}$ (0,0)} bands are identified as well. The review of molecular data focuses primarily on all these detected band systems.

Spectroscopic information is collected for transitions restricted to the lowest rotational quantum states $J = 0 - 5$, which are typically populated in the cold environments investigated in quasar absorption studies \citep[$T \sim 10$ K at $z \sim 2.5$;][]{Srianand2008,Noterdaeme2010}. The CO rotational transitions in the band systems mentioned are described using four molecular parameters: the rest wavelength $\lambda_{i}$, the rotational line oscillator strength $f^{i}_{\jj}$, the natural damping constant $\gamma_{i}$, and the sensitivity coefficient $K_{i}$ to a variation of the proton-to-electron mass ratio. These parameters, specific for each transition, are derived directly from laboratory measurements or via calculational methods. In the following section, methods to derive these molecular parameters are first outlined, and then the values for each electronic band system of CO are presented  in subsequent subsections. Molecular parameters relative to the detected bands, including the undetected A-X$(9 - 0)$ band (see \mbox{Section \ref{subsec:co_bands}}), are listed in order to provide a complete database for future uses in studies of quasar spectra. To avoid any ambiguity, in the following sections the adopted notation and relationships are presented explicitly.

\subsection{Oscillator strengths} \label{subsec:fjj}
In SI units, the absorption oscillator strength of a single rotational transition in a vibronic molecular band is defined as:
\begin{equation}
f_{\jj} = 4 \pi \epsilon_{0} \frac{m_{e} c}{\pi e^{2}}\frac{B^{abs}_{\jj}}{4 \pi} \frac{h c}{\lambda_{\jj}},
\label{eq_def_fjj}
\end{equation}
where $\lambda_{\jj}$ is the transition wavelength, $J'$ refers to the upper level of the transition, $J''$ to the lower level, and the other constants have their traditional meaning. $B^{abs}_{\jj}$ is the Einstein absorption coefficient defined as:
\begin{equation}
B^{abs}_{\jj} = \frac{\lambda_{\jj}^{3} A_{\jj} (2 J' + 1)}{2 h c (2 J'' + 1)},
\label{eq_bcoeff}
\end{equation}
which is connected to the $A_{\jj}$ Einstein coefficient for spontaneous emission:
\begin{equation}
A_{\jj} = \frac{1}{4 \pi \epsilon_{0}} \frac{64 \pi^{4}}{3 h} \frac{1}{\lambda_{\jj}^3} \frac{S_{\jj}}{(2 J' + 1)}.
\label{eq_acoeff}
\end{equation}
$S_{\jj}$ is a line strength factor connected to the squared transition dipole matrix element:
\begin{equation}
S_{\jj} \equiv \sum\limits_{M'} \sum\limits_{M''} |\langle \psi_{J',\,M'} | \overline{\mu} | \psi_{J'',\,M''} \rangle|^{2}
\label{eq_dip_matr_el}
\end{equation}
summed over all magnetic substates \emph{M} of each \emph{J} level, and the operator $\overline{\mu}$ is the electronic transition dipole moment. Using \mbox{Eq. (\ref{eq_bcoeff})} and \mbox{Eq. (\ref{eq_acoeff})}, the line oscillator strength can be rewritten as:
\begin{equation}
f_{\jj} = \frac{8 \pi^{2} m_{e} c}{3 h e^{2}} \frac{S_{\jj}}{\lambda_{\jj} (2 J'' + 1)}.
\label{eq_f_no_coeff}
\end{equation}
In the Born-Oppenheimer approximation, the wavefunction of the molecule can be rewritten in terms of a product of its electronic and nuclear, i.e. vibrational and rotational, components resulting in the relation:
\begin{equation}
S_{\jj} = q_{\vv} |R_{e}(r_{\vv})|^{2} s_{\jj},
\label{eq_boapprox}
\end{equation}
in which $q_{\vv} = |\langle\psi_{\nu'}|\psi_{\nu''}\rangle|^{2}$ is the Franck-Condon factor, $R_{e}(r_{\vv})$ is the electric dipole moment function, $r_{\vv}$ is the vibrationally averaged internuclear distance, and $s_{\jj}$ is the H\"onl-London factor for spin-allowed transitions. Inserting \mbox{Eq. (\ref{eq_boapprox})} in \mbox{Eq. (\ref{eq_f_no_coeff})} gives the final expression for the oscillator strength of a rotational line in a rovibronic molecular band:
\begin{equation}
f_{\jj} = \frac{8 \pi^{2} m_{e} c}{3 h e^{2}} q_{\vv} |R_{e}(r_{\vv})|^{2} \frac{s_{\jj}}{\lambda_{\jj} (2 J'' + 1)}.
\label{eq_fjj_nofvv}
\end{equation}

In experimental spectroscopy, it is not always possible to separately resolve individual rotational transitions for each \emph{J} ground state level. Often an entire set of transitions associated with vibrational levels $\nuprime$ and $\nu\,''$, hereafter referred to as a vibrational band, is addressed. It is possible to define an oscillator strength for a vibrational band, in analogy to \mbox{Eq. (\ref{eq_fjj_nofvv})}:
\begin{equation}
f_{\vv} = \frac{8 \pi^{2} m_{e} c}{3 h e^{2}} q_{\vv} |R_{e}(r_{\vv})|^{2} \frac{1}{\lambda_{\vv}} \delta_{\Sigma,\Pi},
\label{eq_fvv}
\end{equation}
where $\lambda_{\vv}$ is the wavelength of the band origin, and $\delta_{\Sigma,\Pi} = \{1, 2\}$ for $\Sigma$ and $\Pi$ excited states respectively. This band oscillator strength is equivalent to a sum over all single line oscillator strengths $f_{\jj}$. Hence there is a relation between the band oscillator strength and the line oscillator strength:
\begin{equation}
f_{\jj} = f_{\vv} \frac{\lambda_{\vv}}{\lambda_{\jj}} \frac{s_{\jj}}{(2J +1)} \frac{1}{\delta_{\Sigma,\Pi}}.
\label{eq_fjj}
\end{equation}
Note that the ratio between the wavelengths of the vibrational band origin $\lambda_{\vv}$ and the individual rotational lines $\lambda_{\jj}$ is \mbox{$\simeq 1$}, because the spread in the wavelengths of each vibrational band is limited and the band origin wavelength can be considered as the average wavelength of the band.

The above analysis and \mbox{Eq. (\ref{eq_fjj})} is valid if no perturbations occur in the electronic structure. In case of a perturbation, a transition to a perturber state, which could be forbidden, `borrows' intensity from an allowed electronic transition. Hence the effective band oscillator strength will be divided between the allowed and forbidden transitions. The actual wavefunction $\Psi_{i}$ is mixture of the zero-order wavefunctions $\Psi^{0}_{i}$'s, the latter describing the system in the absence of the non-diagonal terms in the interaction Hamiltonian. $\Psi_{i}$ can be expressed as a linear superposition
\begin{equation}
\Psi_i = \alpha_{i} \Psi_{i}^{0} + \sum_{j} \alpha_{j} \Psi^{0}_{j},
\label{eq_mix_pert}
\end{equation}
where $\alpha$ are the mixing coefficients obtained in a deperturbation analysis \citep[e.g.][]{Niu2013}, that are normalized at \mbox{$|\alpha_{i}|^{2} + \sum |\alpha_{j}|^{2} = 1$}. Note that the \emph{i}-th state is normally assigned to the state with the greatest $\alpha$, i.e. the dominant electronic wavefunction character, and that the summation holds for the case where multiple perturbing states are involved. As will be discussed below for d-X transitions, dipole-forbidden transitions can be observed if an interaction exists with levels that have dipole-allowed transitions, e.g. the A($\nu = 1$) and d($\nu = 5$) levels. In this case, the oscillator strengths $f_{\textrm{\scriptsize d5}}$ can be expressed as:
\begin{equation}
f_{\textrm{\scriptsize d5}} = f^0_{\textrm{\scriptsize A1}} |\alpha_{\textrm{\scriptsize d5}}|^2,
\label{eq_fjj_pert}
\end{equation}
in terms of the dipole-allowed deperturbed oscillator strengths $f^0_{\textrm{\scriptsize A1}}$ of the A-X transitions, for levels with d(v=5) character given by the mixing coefficient $|\alpha_{\textrm{\scriptsize d5}}|^2$. This is a general effect that occurs in the presence of perturbations, where the \emph{intensity borrowing} phenomenon transfers a fraction of the allowed transition oscillator strength to those of forbidden lines. 

\subsection{Damping constants} \label{subsec:gamma}
The natural damping parameter $\gamma_{\nuprime}$ can directly be obtained from experiments through measurement of the excited lifetime $\tau_{\nuprime}$ which, in absence of collisions, predissociation and autoionization, can be expressed in terms of the sum of the Einstein coefficients for spontaneous emission to all the ground state vibrational and rotational levels:
\begin{equation}
\gamma_{\nuprime J\,'} = \frac{1}{\tau_{\nuprime J\,'}} = \sum\limits_{\nu\,'',\,J\,''} A_{\nuprime J\,' \nu\,'' J\,''},
\label{eq_gamma_def}
\end{equation}
which, considering the Born-Oppenheimer approximation and substituting \mbox{Eq. (\ref{eq_acoeff})} and \mbox{Eq. (\ref{eq_boapprox})}, becomes:
\begin{equation}
\gamma_{\nuprime J\,'} = \frac{1}{4 \pi \epsilon_{0}} \frac{64 \pi^{2}}{3 h} \delta_{\Sigma,\,\Pi} \sum\limits_{\nu\,'',\,J\,''} \frac{1}{\lambda_{\jj}^{3}} q_{\vv} |R_{e}(r_{\vv})|^{2} \frac{s_{\jj}}{(2 J' + 1)}.
\label{eq_gamma_ext}
\end{equation}

In case of perturbations, the lifetime is no longer a constant across the \emph{J} levels because of the local interactions between the short-lived perturbed state and the usually long-lived perturbing states. In this case, the damping parameter is given by:
\begin{equation}
\gamma_{\nuprime J\,',\,i} = |\alpha_{i}|^{2} \gamma^{0}_{\nuprime J\,',\,i} + \sum\limits_{j} |\alpha_{j}|^{2} \gamma^{0}_{\nuprime J\,',\,j},
\label{eq_gamma_pert}
\end{equation}
in which $\alpha_{i}$ and $\alpha_{j}$ are the mixing coefficients as in \mbox{Eq. (\ref{eq_mix_pert})}, while $\gamma^{0}_{\nuprime J\,',\,i}$ and $\gamma^{0}_{\nuprime J\,',\,j}$ are the damping parameters of each pure state. Note that the state with the largest $|\alpha|^{2}$ is indicated with the subscript \emph{i} and it is referred to as the `perturbed state', while the other \emph{j} states are referred to as the `perturbing states'.

In cases where the decay is not purely radiative, but includes rates associated with predissociation, the damping rates are associated with the natural lifetimes according to:
\begin{equation}
\gamma_{\nuprime J\,'} = \frac{1}{\tau_{\nuprime J\,'}} = A^{rad}_{\nuprime J\,'} + A^{pred}_{\nuprime J\,'}.
\label{eq_gamma_prediss}
\end{equation}

\subsection{Sensitivity coefficients} \label{subsubsec:ki}
One ingredient in a search for a varying proton-to-electron mass ratio based on spectral lines in molecules is an assessment of the sensitivity coefficients $K_{i}$, which are defined as~\citep{Ubachs2007}:
\begin{equation}
K_{i} = \frac{\mathrm d \ln\lambda_{i}}{\mathrm d \ln\mu}.
\label{ki_pure}
\end{equation}
\cite{Salumbides2012} calculated the sensitivity coefficients for the CO A-X bands up to $\nuprime = 10$, starting from a Dunham expansion of the rovibrational level energies. 
The calculation of the sensitivity coefficients exploits the known mass-dependence of the Dunham parameters, which are obtained in the semi-empirical modelling of the rovibrational level energies. While this procedure is straightforward for unperturbed states, the presence of local interactions introduces a difficulty, since a state is in fact a mixture of two or more states, which in general have energies with different mass-dependencies. To account for the effects of local interactions, $\mu$-sensitivity coefficients $K^{0}$ are first calculated in the \emph{idealized} case when the interaction between bands is zero, i.e. using the diagonal terms in the interaction matrix obtained in a deperturbation analysis. Subsequently, the \emph{true} \emph{K}-coefficients that include the effect of perturbations are recovered, to first-order, by adopting the wavefunction admixture coefficients as weights in a relation similar to that of \mbox{Eq. (\ref{eq_mix_pert})}:
\begin{equation}
K_{i} = |\alpha_{i}|^{2} K^{0}_{i} + \sum_{j} |\alpha_{j}|^{2} K^{0}_{j},
\label{ki_pert}
\end{equation}
where $\alpha$ and $K^{0}$ values are obtained using results from the deperturbation analysis.

\subsection{The A-X system} \label{subsec:axsyst}
The excited electronic structure of the CO molecule is a textbook example of perturbations involving a large number of states of singlet and triplet character. Many of the details of these perturbations had been unravelled by \cite{Field1972} and \cite{Lefloch1987}. For the case of quasar absorption analysis of CO in particular the perturbation between the  \mbox{A-X$(1 - 0)$} and \mbox{d-X$(5 - 0)$} vibrational bands is of relevance, because the interaction is strongest at the low \emph{J} levels that are populated in cold clouds. This specific case has been re-analyzed by \cite{Niu2013}.

Accurate rest wavelengths are compiled from \cite{Salumbides2012} and \cite{Niu2013,Niu2015}, who measured CO transitions in five bands, detecting \mbox{$\sim 200$ lines} per each band up to the rotational level with \mbox{$J = 50$}. These laboratory wavelengths were determined using two independent studies: a vacuum ultraviolet Fourier-transform (VUV-FT) and a two-photon Doppler-free laser-based excitation experiment. Starting from these measurements, \cite{Niu2013} developed a semi-empirical model, including all the perturbing states relative to the five observed A-X$(\nuprime - 0)$ bands for $\nuprime = 0 - 4$. These studies delivered CO transitions wavelengths, including A-X$(\nuprime - 0)$ bands for $\nuprime = 5 - 9$, with an accuracy better than \mbox{$\Delta\lambda/\lambda = 3 \times 10^{-7}$} and can be considered exact for comparisons with the observed CO bands in quasar spectra. Laboratory wavelengths for the A-X system for \mbox{$J=0 - 5$} ground state levels are listed in \mbox{Table \ref{wl_ax}}.

Band oscillator strength $f_{\jj}$ values for the A-X system were measured using a variety of different techniques. \cite{Eidelsberg1992} used optical absorption, while \cite{Chan1993} and \cite{Zhong1998} determined the band oscillator strengths using electron scattering. An additional set of $f_{\vv}$ values was calculated via \mbox{Eq. (\ref{eq_fvv})}, using the Franck-Condon factors reported by \cite{Eidelsberg1992}, while the dipole moment was calculated starting from the lifetime measurements of \cite{Field1983}. The linear dipole-moment function \mbox{$R_{e}(r_{\vv}) = 7.48(34)[1 - 0.683(7)r_{\vv}]$}~\citep[the dipole moment is in Debyes and $r_{\vv}$ is in \AA,][]{Field1983} was adopted, rather than the extended parabolic one presented by \cite{DeLeon1989}, because the latter does not represent well the low $\nu$ levels considered in this work. This results in four sets of band oscillator strength values, with an accuracy of \mbox{$\sim 10\%$} or better, which agree with each other within their combined error. While \cite{Morton1994} relied entirely on values derived by \cite{Chan1993}, presently the four sets are averaged together to obtain a value for the band oscillator strength which was used to calculate the $f_{\jj}$ values for each transition based on \mbox{Eq. (\ref{eq_fjj})}. For the single case of the A-X$(1 - 0)$ band, which is perturbed by the d-X$(5 - 0)$ band, the state mixing analysis of \mbox{Eq. (\ref{eq_fjj_pert})} was invoked to determine the $f_{\jj}$ values. The mixing coefficients were obtained from the revised perturbation analysis performed by \cite{Niu2013}. The line oscillator strengths for the A-X system are reported in \mbox{Table \ref{fjj_ax}}, and the sensitivity coefficients from \cite{Salumbides2012} are listed in \mbox{Table \ref{ki_ax}}. 

\cite{Field1983} measured the lifetimes for the levels $\nuprime = 0 - 7$ with an accuracy of \mbox{$\sim 1\%$}, and these values were used to estimate lifetimes for states $\nuprime = 8 - 9$. The damping parameters relative to the A-X system were calculated using \mbox{Eq. (\ref{eq_gamma_def})} and \mbox{Eq. (\ref{eq_gamma_pert})}. Note that the value of $\gamma_{\nuprime}$ for the level A$^{1} \Pi$, $\nuprime = 1$ includes the effect of the perturbation by the longer lived d$^{3} \Delta$, $\nuprime = 5$. The other A$^{1} \Pi$, $\nuprime$ states are mainly perturbed at high rotational states, hence they were considered unperturbed for the relevant low-\emph{J} levels during this work. The damping parameters $\gamma_{\nuprime}$ of the A-X bands are presented in \mbox{Table \ref{gamma_ax}}.
\begin{deluxetable}{cllllll}
\tabletypesize{\scriptsize}
\tablecaption{Laboratory wavelengths for A-X$ (\nuprime-0)$ bands of CO. \label{wl_ax}}
\tablewidth{0pt}
\tablehead{
\colhead{\emph{J"}} & \multicolumn{3}{c}{Wavelength [\AA]} & \multicolumn{3}{c}{Wavelength [\AA]} \\
 & \colhead{R} & \colhead{Q} & \colhead{P} & \colhead{R} & \colhead{Q} & \colhead{P}
}
\startdata
 & \multicolumn{3}{c}{\textbf{(0-0)}\tablenotemark{a}} & \multicolumn{3}{c}{\textbf{(1-0)}\tablenotemark{a}} \\
0 & 1544.44965(5)\tablenotemark{*} & \multicolumn{1}{c}{-} & \multicolumn{1}{c}{-} & 1509.74781(5)\tablenotemark{*} & \multicolumn{1}{c}{-} & \multicolumn{1}{c}{-} \\
1 & 1544.38952(5)\tablenotemark{*} & 1544.54133(5)\tablenotemark{*} & \multicolumn{1}{c}{-} & 1509.69595(5)\tablenotemark{*} & 1509.83545(5)\tablenotemark{*} & \multicolumn{1}{c}{-} \\
2 & 1544.34528(5)\tablenotemark{*} & 1544.57263(5)\tablenotemark{*} & 1544.72485(5)\tablenotemark{*} & 1509.66147(5)\tablenotemark{*} & 1509.87124(5)\tablenotemark{*} & 1510.01077(5)\tablenotemark{*} \\
3 & 1544.31734(24)\tablenotemark{\dag} & 1544.61950(5)\tablenotemark{*} & 1544.84819(5)\tablenotemark{*} & 1509.64364(5)\tablenotemark{*} & 1509.92437(5)\tablenotemark{*} & 1510.13424(5)\tablenotemark{*} \\
4 & 1544.30586(24)\tablenotemark{\dag} & 1544.68226(5)\tablenotemark{*} & 1544.98742(5)\tablenotemark{*} & 1509.64196(23)\tablenotemark{\dag} & 1509.99415(5)\tablenotemark{*} & 1510.27510(5)\tablenotemark{*} \\
5 & 1544.31235(24)\tablenotemark{\dag} & 1544.76096(5)\tablenotemark{*} & 1545.14298(24)\tablenotemark{\dag} & 1509.65592(23)\tablenotemark{\dag} & 1510.08008(5)\tablenotemark{*} & 1510.43261(5)\tablenotemark{*} \\
 & \multicolumn{3}{c}{\textbf{(2-0)}\tablenotemark{b}} & \multicolumn{3}{c}{\textbf{(3-0)}\tablenotemark{b}} \\
0 & 1477.56549(4)\tablenotemark{*} & \multicolumn{1}{c}{-} & \multicolumn{1}{c}{-} & 1447.35311(4)\tablenotemark{*} & \multicolumn{1}{c}{-} & \multicolumn{1}{c}{-} \\
1 & 1477.51338(4)\tablenotemark{*} & 1477.64944(4)\tablenotemark{*} & \multicolumn{1}{c}{-} & 1447.30514(4)\tablenotemark{*} & 1447.43368(4)\tablenotemark{*} & \multicolumn{1}{c}{-} \\
2 & 1477.47721(4)\tablenotemark{*} & 1477.68137(4)\tablenotemark{*} & 1477.81737(4)\tablenotemark{*} & 1447.27343(4)\tablenotemark{*} & 1447.46624(4)\tablenotemark{*} & 1447.59479(4)\tablenotemark{*} \\
3 & 1477.45705(4)\tablenotemark{*} & 1477.72920(4)\tablenotemark{*} & 1477.93318(4)\tablenotemark{*} & 1447.25798(4)\tablenotemark{*} & 1447.51507(4)\tablenotemark{*} & 1447.70795(4)\tablenotemark{*} \\
4 & 1477.45279(4)\tablenotemark{*} & 1477.79304(4)\tablenotemark{*} & 1478.06494(4)\tablenotemark{*} & 1447.25885(4)\tablenotemark{*} & 1447.58021(4)\tablenotemark{*} & 1447.83737(4)\tablenotemark{*} \\
5 & 1477.46449(4)\tablenotemark{*} & 1477.87294(4)\tablenotemark{*} & 1478.21273(4)\tablenotemark{*} & 1447.27595(4)\tablenotemark{*} & 1447.66165(4)\tablenotemark{*} & 1447.98308(4)\tablenotemark{*} \\
 & \multicolumn{3}{c}{\textbf{(4-0)}\tablenotemark{b}} & \multicolumn{3}{c}{\textbf{(5-0)}\tablenotemark{c}} \\
0 & 1419.04491(4)\tablenotemark{*} & \multicolumn{1}{c}{-} & \multicolumn{1}{c}{-} & 1392.52551(20)\tablenotemark{\dag} & \multicolumn{1}{c}{-} & \multicolumn{1}{c}{-} \\
1 & 1419.00170(4)\tablenotemark{*} & 1419.12198(4)\tablenotemark{*} & \multicolumn{1}{c}{-} & 1392.48517(20)\tablenotemark{\dag} & 1392.60017(20)\tablenotemark{\dag} & \multicolumn{1}{c}{-} \\
2 & 1418.97534(4)\tablenotemark{*} & 1419.15561(4)\tablenotemark{*} & 1419.27723(4)\tablenotemark{*} & 1392.46133(20)\tablenotemark{\dag} & 1392.63391(20)\tablenotemark{\dag} & 1392.74951(20)\tablenotemark{\dag} \\
3 & 1418.96575(4)\tablenotemark{*} & 1419.20618(4)\tablenotemark{*} & 1419.38890(4)\tablenotemark{*} & 1392.45435(20)\tablenotemark{\dag} & 1392.68473(20)\tablenotemark{\dag} & 1392.85776(20)\tablenotemark{\dag} \\
4 & 1418.97285(20)\tablenotemark{\dag} & 1419.27367(4)\tablenotemark{*} & 1419.51744(4)\tablenotemark{*} & 1392.46365(20)\tablenotemark{\dag} & 1392.75203(20)\tablenotemark{\dag} & 1392.98290(20)\tablenotemark{\dag} \\
5 & 1418.99625(20)\tablenotemark{\dag} & 1419.35858(20)\tablenotemark{\dag} & 1419.66277(4)\tablenotemark{*} & 1392.49041(20)\tablenotemark{\dag} & 1392.83642(20)\tablenotemark{\dag} & 1393.12476(20)\tablenotemark{\dag} \\
 & \multicolumn{3}{c}{\textbf{(6-0)}\tablenotemark{c}} & \multicolumn{3}{c}{\textbf{(7-0)}\tablenotemark{c}} \\
0 & 1367.62386(20)\tablenotemark{\dag} & \multicolumn{1}{c}{-} & \multicolumn{1}{c}{-} & 1344.18593(20)\tablenotemark{\dag} & \multicolumn{1}{c}{-} & \multicolumn{1}{c}{-} \\
1 & 1367.58645(20)\tablenotemark{\dag} & 1367.69587(20)\tablenotemark{\dag} & \multicolumn{1}{c}{-} & 1344.15142(20)\tablenotemark{\dag} & 1344.25550(20)\tablenotemark{\dag} & \multicolumn{1}{c}{-} \\
2 & 1367.56681(20)\tablenotemark{\dag} & 1367.73029(20)\tablenotemark{\dag} & 1367.83936(20)\tablenotemark{\dag} & 1344.13426(20)\tablenotemark{\dag} & 1344.29056(20)\tablenotemark{\dag} & 1344.39411(20)\tablenotemark{\dag} \\
3 & 1367.56494(20)\tablenotemark{\dag} & 1367.78248(20)\tablenotemark{\dag} & 1367.94620(20)\tablenotemark{\dag} & 1344.13480(20)\tablenotemark{\dag} & 1344.34278(20)\tablenotemark{\dag} & 1344.49877(20)\tablenotemark{\dag} \\
4 & 1367.58065(20)\tablenotemark{\dag} & 1367.85264(20)\tablenotemark{\dag} & 1368.07047(20)\tablenotemark{\dag} & 1344.15233(20)\tablenotemark{\dag} & 1344.41273(20)\tablenotemark{\dag} & 1344.62098(20)\tablenotemark{\dag} \\
5 & 1367.61544(20)\tablenotemark{\dag} & 1367.94040(20)\tablenotemark{\dag} & 1368.21216(20)\tablenotemark{\dag} & 1344.18756(20)\tablenotemark{\dag} & 1344.49985(20)\tablenotemark{\dag} & 1344.76003(20)\tablenotemark{\dag} \\
 & \multicolumn{3}{c}{\textbf{(8-0)}\tablenotemark{c}} & \multicolumn{3}{c}{\textbf{(9-0)}\tablenotemark{c}} \\
0 & 1322.15058(20)\tablenotemark{\dag} & \multicolumn{1}{c}{-} & \multicolumn{1}{c}{-} & 1301.40162(35)\tablenotemark{\dag} & \multicolumn{1}{c}{-} & \multicolumn{1}{c}{-} \\
1 & 1322.11912(20)\tablenotemark{\dag} & 1322.21754(20)\tablenotemark{\dag} & \multicolumn{1}{c}{-} & 1301.37300(35)\tablenotemark{\dag} & 1301.46598(35)\tablenotemark{\dag} & \multicolumn{1}{c}{-} \\
2 & 1322.10531(20)\tablenotemark{\dag} & 1322.25355(20)\tablenotemark{\dag} & 1322.35252(20)\tablenotemark{\dag} & 1301.36148(35)\tablenotemark{\dag} & 1301.50274(35)\tablenotemark{\dag} & 1301.59659(35)\tablenotemark{\dag} \\
3 & 1322.10898(20)\tablenotemark{\dag} & 1322.30671(20)\tablenotemark{\dag} & 1322.45534(20)\tablenotemark{\dag} & 1301.36894(35)\tablenotemark{\dag} & 1301.55746(35)\tablenotemark{\dag} & 1301.69740(35)\tablenotemark{\dag} \\
4 & 1322.13118(20)\tablenotemark{\dag} & 1322.37822(20)\tablenotemark{\dag} & 1322.57585(20)\tablenotemark{\dag} & 1301.39349(35)\tablenotemark{\dag} & 1301.62946(35)\tablenotemark{\dag} & 1301.81873(35)\tablenotemark{\dag} \\
5 & 1322.17069(20)\tablenotemark{\dag} & 1322.46724(20)\tablenotemark{\dag} & 1322.71423(20)\tablenotemark{\dag} & 1301.43702(35)\tablenotemark{\dag} & 1301.71892(35)\tablenotemark{\dag} & 1301.95399(35)\tablenotemark{\dag} \\
\enddata
\tablenotetext{a}{From \cite{Niu2013}.}
\tablenotetext{b}{From \cite{Niu2015}.}
\tablenotetext{c}{From \cite{Salumbides2012}.}
\tablenotetext{*}{Derived from the laser study.}
\tablenotetext{\dag}{Derived from the VUV-FTS study.}
\end{deluxetable}
 
\begin{deluxetable}{ccccccc}
\tabletypesize{\scriptsize}
\tablecaption{Line oscillator strengths for A-X$ (\nuprime-0)$ bands of CO. \label{fjj_ax}}
\tablewidth{0pt}
\tablehead{
\colhead{\emph{J"}} & \multicolumn{3}{c}{f$_{\jj}$} & \multicolumn{3}{c}{f$_{\jj}$} \\
 & \colhead{R} & \colhead{Q} & \colhead{P} & \colhead{R} & \colhead{Q} & \colhead{P}
}
\startdata
 & \multicolumn{3}{c}{\textbf{(0-0)}} & \multicolumn{3}{c}{\textbf{(1-0)}} \\
0 & 1.582E-02 & - & - & 2.835E-02 & - & - \\
1 & 7.893E-03 & 7.908E-03 & - & 1.439E-02 & 1.417E-02 & - \\
2 & 6.294E-03 & 7.893E-03 & 1.582E-03 & 1.174E-02 & 1.439E-02 & 2.835E-03 \\
3 & 5.590E-03 & 7.868E-03 & 2.255E-03 & 1.072E-02 & 1.468E-02 & 4.111E-03 \\
4 & 5.168E-03 & 7.826E-03 & 2.623E-03 & 1.023E-02 & 1.501E-02 & 4.893E-03 \\
5 & 4.847E-03 & 7.753E-03 & 2.846E-03 & 9.969E-03 & 1.535E-02 & 5.458E-03 \\
 & \multicolumn{3}{c}{\textbf{(2-0)}} & \multicolumn{3}{c}{\textbf{(3-0)}} \\
0 & 4.151E-02 & - & -                 & 3.651E-02 & - & - \\
1 & 2.075E-02 & 2.075E-02 & -	      & 1.825E-02 & 1.825E-02 & - \\	      
2 & 1.660E-02 & 2.075E-02 & 4.151E-03 & 1.460E-02 & 1.825E-02 & 3.651E-03 \\
3 & 1.482E-02 & 2.075E-02 & 5.930E-03 & 1.304E-02 & 1.825E-02 & 5.216E-03 \\
4 & 1.384E-02 & 2.075E-02 & 6.918E-03 & 1.217E-02 & 1.826E-02 & 6.085E-03 \\
5 & 1.321E-02 & 2.075E-02 & 7.547E-03 & 1.162E-02 & 1.826E-02 & 6.639E-03 \\
 & \multicolumn{3}{c}{\textbf{(4-0)}} & \multicolumn{3}{c}{\textbf{(5-0)}} \\
0 & 2.448E-02 & - & -                 & 1.582E-02 & - & - \\
1 & 1.229E-02 & 1.224E-02 & -	      & 7.912E-03 & 7.912E-03 & - \\	      
2 & 9.888E-03 & 1.229E-02 & 2.448E-03 & 6.330E-03 & 7.912E-03 & 1.582E-03 \\
3 & 8.876E-03 & 1.236E-02 & 3.513E-03 & 5.652E-03 & 7.912E-03 & 2.261E-03 \\
4 & 8.448E-03 & 1.243E-02 & 4.120E-03 & 5.275E-03 & 7.912E-03 & 2.637E-03 \\
5 & 7.977E-03 & 1.249E-02 & 4.519E-03 & 5.035E-03 & 7.912E-03 & 2.877E-03 \\
 & \multicolumn{3}{c}{\textbf{(6-0)}} & \multicolumn{3}{c}{\textbf{(7-0)}} \\
0 & 9.501E-03 & - & -                 & 5.354E-03 & - & - \\
1 & 4.746E-03 & 4.750E-03 & -	      & 2.677E-03 & 2.677E-03 & - \\	      
2 & 3.789E-03 & 4.746E-03 & 9.501E-04 & 2.142E-03 & 2.677E-03 & 5.354E-04 \\
3 & 3.371E-03 & 4.736E-03 & 1.356E-03 & 1.912E-03 & 2.677E-03 & 7.649E-04 \\
4 & 3.121E-03 & 4.720E-03 & 1.579E-03 & 1.785E-03 & 2.677E-03 & 8.923E-04 \\
5 & 2.902E-03 & 4.682E-03 & 1.716E-03 & 1.704E-03 & 2.677E-03 & 9.735E-04 \\
 & \multicolumn{3}{c}{\textbf{(8-0)}} & \multicolumn{3}{c}{\textbf{(9-0)}} \\
0 & 2.636E-03 & - & -                 & 1.294E-03 & - & - \\
1 & 1.318E-03 & 1.318E-03 & -	      & 6.470E-04 & 6.470E-04 & - \\	      
2 & 1.054E-03 & 1.318E-03 & 2.636E-04 & 5.176E-04 & 6.470E-04 & 1.294E-04 \\
3 & 9.414E-04 & 1.318E-03 & 3.766E-04 & 4.621E-04 & 6.470E-04 & 1.849E-04 \\
4 & 8.787E-04 & 1.318E-03 & 4.393E-04 & 4.313E-04 & 6.470E-04 & 2.157E-04 \\
5 & 8.387E-04 & 1.318E-03 & 4.793E-04 & 4.117E-04 & 6.470E-04 & 2.353E-04 \\
\enddata
\tablecomments{Uncertainties on the oscillator strengths are better than \mbox{$\sim10\%$}.}
\end{deluxetable}

\begin{deluxetable}{ccccccc}
\tabletypesize{\scriptsize}
\tablecaption{Sensitivity coefficients for A-X$ (\nuprime-0)$ bands of CO. \label{ki_ax}}
\tablewidth{0pt}
\tablehead{
\colhead{\emph{J"}} & \multicolumn{3}{c}{$K_{i}$} & \multicolumn{3}{c}{$K_{i}$} \\
 & \colhead{R} & \colhead{Q} & \colhead{P} & \colhead{R} & \colhead{Q} & \colhead{P}
 }
\startdata
 & \multicolumn{3}{c}{\textbf{(0-0)}} & \multicolumn{3}{c}{\textbf{(1-0)}} \\
0 & -0.00232 & - & - & 0.01312	& - & - \\
1 & -0.00227 & -0.00237 & - & 0.01280 & 0.01306 & - \\
2 & -0.00223 & -0.00238 & -0.00249 & 0.01235 & 0.01269 & 0.01294 \\
3 & -0.00219 & -0.00239 & -0.00257 & 0.01183 & 0.01218 & 0.01251 \\
4 & -0.00216 & -0.00240 & -0.00264 & 0.01129 & 0.01160 & 0.01195 \\
5 & -0.00212 & -0.00239 & -0.00273 & 0.01077 & 0.01100 & 0.01132 \\
 & \multicolumn{3}{c}{\textbf{(2-0)}} & \multicolumn{3}{c}{\textbf{(3-0)}} \\
0 & 0.01850 & - & - & 0.02756 & - & - \\
1 & 0.01853 & 0.01844 & - & 0.02759 & 0.02750 & - \\
2 & 0.01855 & 0.01842 & 0.01833 & 0.02761 & 0.02748 & 0.02740 \\
3 & 0.01856 & 0.01838 & 0.01825 & 0.02761 & 0.02744 & 0.02732 \\
4 & 0.01856 & 0.01834 & 0.01816 & 0.02761 & 0.02739 & 0.02723 \\
5 & 0.01854 & 0.01828 & 0.01806 & 0.02759 & 0.02734 & 0.02712 \\
 & \multicolumn{3}{c}{\textbf{(4-0)}} & \multicolumn{3}{c}{\textbf{(5-0)}} \\
0 & 0.03784 & - & - & 0.04329 & - & - \\
1 & 0.03783 & 0.03697 & - & 0.04331 & 0.04324 & - \\
2 & 0.03771 & 0.03680 & 0.03695 & 0.04332 & 0.04321 & 0.04314 \\
3 & 0.03749 & 0.03659 & 0.03685 & 0.04332 & 0.04317 & 0.04306 \\
4 & 0.03720 & 0.03636 & 0.03667 & 0.04331 & 0.04312 & 0.04297 \\
5 & 0.03690 & 0.03614 & 0.03643 & 0.04328 & 0.04305 & 0.04286 \\
 & \multicolumn{3}{c}{\textbf{(6-0)}} & \multicolumn{3}{c}{\textbf{(7-0)}} \\
0 & 0.05039 & - & - & 0.05612 & - & - \\
1 & 0.05044 & 0.05034 & - & 0.05614 & 0.05607 & - \\
2 & 0.05051 & 0.05034 & 0.05024 & 0.05615 & 0.05605 & 0.05598 \\
3 & 0.05061 & 0.05034 & 0.05019 & 0.05614 & 0.05600 & 0.05590 \\
4 & 0.05066 & 0.05037 & 0.05016 & 0.05611 & 0.05594 & 0.05581 \\
5 & 0.05124 & 0.05048 & 0.05016 & 0.05607 & 0.05587 & 0.05570 \\
 & \multicolumn{3}{c}{\textbf{(8-0)}} & \multicolumn{3}{c}{\textbf{(9-0)}} \\
0 & 0.06162 & - & - & 0.06646 & - & - \\
1 & 0.06163 & 0.06157 & - & 0.06647 & 0.06641 & - \\
2 & 0.06169 & 0.06154 & 0.06147 & 0.06647 & 0.06638 & 0.06632 \\
3 & 0.06162 & 0.06149 & 0.06139 & 0.06646 & 0.06633 & 0.06624 \\
4 & 0.06159 & 0.06143 & 0.06136 & 0.06642 & 0.06627 & 0.06615 \\
5 & 0.06154 & 0.06135 & 0.06119 & 0.06638 & 0.06619 & 0.06604 \\
\enddata
\tablecomments{Uncertainties on the sensitivity coefficients are better than \mbox{$\sim1\%$}.}
\end{deluxetable}

\begin{deluxetable}{cc}
\tablecaption{Natural damping constants for A-X$(\nuprime - 0)$ bands of CO. \label{gamma_ax}}
\tablewidth{0pt}
\tablehead{
\colhead{$\nuprime$} & $\gamma_{\nuprime}$\tablenotemark{a} [$\times 10^{8}$ s$^{-1}$]}
\startdata
0 & 1.00 \\
1 & 0.96\tablenotemark{b} \\
2 & 1.03 \\
3 & 1.04 \\
4 & 1.06 \\
5 & 1.09 \\
6 & 1.10 \\
7 & 1.12 \\
8 & 1.12\tablenotemark{c} \\
9 & 1.12\tablenotemark{c} \\
\enddata
\tablenotetext{a}{From \cite{Field1983}.}
\tablenotetext{b}{Including the effect of perturbation.}
\tablenotetext{c}{Values estimated.}
\end{deluxetable}

\subsection{The d-X system} \label{subsec:dxsyst}
The laboratory wavelengths of the d-X$(5 - 0)$ band up to levels with \mbox{$J= 10$} were partially listed by \cite{Niu2013}. Wavelengths for the transitions that were not directly measured in the VUV-FT or in the laser studies, were calculated using the semi-empirical model developed by \cite{Niu2013}, since it is more accurate and consistent than the other laboratory measurements of the d-X band present in literature.

The d-X band is a spin-forbidden transition which borrows its intensity from the interaction between the A$^{1} \Pi$ and d$^{3} \Delta$ states. Values for the line oscillator strengths $f_{\jj}$  are obtained via state mixing as governed by \mbox{Eq. (\ref{eq_fjj_pert})}, where the mixing coefficients are obtained from the perturbation analysis of \cite{Niu2013}.

A similar procedure was used to derive the sensitivity coefficients of this band, which were calculated according to \mbox{Eq. (\ref{ki_pert})}, using as $K^{0}_{i}$  the sensitivity coefficients of the unperturbed d-X transitions and as $K^{0}_{j}$ the unperturbed coefficients of the A-X$(1-0)$ band transitions. In other words, the sensitivity coefficients for the d-X transitions were calculated considering the \mbox{A-X$(1-0)$} band as the perturber of the \mbox{d-X$ (5-0)$} band. 

Lifetimes for d$^{3} \Delta$ state with $\nuprime = 1 - 16$ were measured by \cite{Vansprang1977} and they found lifetimes depending on vibrational level between 7.3 and \mbox{2.9 $\mu$s} increasing with vibrational quantum number. These measured lifetimes are averages of the lifetimes of rotational levels in a vibrational manifold. As a consequence of the state mixing, a \emph{J} dependent damping parameter was obtained for each rotational state in d$^{3} \Delta$, $\nuprime = 5$ following \mbox{Eq.(\ref{eq_gamma_pert})} and the mixing coefficients from the perturbation model by \cite{Niu2015}.

\begin{deluxetable}{clllccccccc}
\rotate
\tablecaption{Molecular parameters for the d-X$ (5-0)$ band. \label{dx_band}}
\tablewidth{0pt}
\tablehead{
\colhead{\emph{J"}} & \multicolumn{3}{c}{Wavelength [\AA]} & \multicolumn{3}{c}{f$_{\jj}$} & \multicolumn{3}{c}{\emph{K$_{i}$}} & $\gamma_{\nuprime}$\tablenotemark{d} [$\times 10^{8}$ s$^{-1}$] \\
 & \colhead{R} & \colhead{Q} & \colhead{P} & \colhead{R} & \colhead{Q} & \colhead{P} & \colhead{R} & \colhead{Q} & \colhead{P} & 
}
\startdata
0 & 1510.34131(20)\tablenotemark{b} & - & - & 5.70E-03 & - & - & 0.031298 & - & - & 0.18 \\
1 & 1510.30590(5)\tablenotemark{c} & 1510.42893(5)\tablenotemark{c} & - & 2.64E-03 & 2.85E-03 & - & 0.031660 & 0.031298 & - & 0.16 \\
2 & 1510.29756(5)\tablenotemark{c} & 1510.48138(5)\tablenotemark{a} & 1510.60448(20)\tablenotemark{b} & 1.88E-03 & 2.64E-03 & 5.70E-04 & 0.032123 & 0.031634 & 0.031271 & 0.14 \\
3 & 1510.31694(5)\tablenotemark{c} & 1510.56071(5)\tablenotemark{a} & 1510.74454(5)\tablenotemark{c} & 1.45E-03 & 2.35E-03 & 7.53E-04 & 0.032628 & 0.032072 & 0.031581 & 0.13 \\
4 & 1510.36485(5)\tablenotemark{a} & 1510.66783(5)\tablenotemark{a} & 1510.91169(5)\tablenotemark{c} & 1.13E-03 & 2.02E-03 & 7.84E-04 & 0.033123 & 0.032553 & 0.031994 & 0.10 \\
5 & 1510.44207(5)\tablenotemark{a} & 1510.80354(5)\tablenotemark{c} & 1511.10612(20)\tablenotemark{b} & 8.76E-04 & 1.69E-03 & 7.36E-04 & 0.033562 & 0.033027 & 0.032453 & 0.09 \\
\enddata
\tablecomments{Uncertainties in sensitivity coefficients estimated to be better than $\sim 1\%$.}
\tablenotetext{a}{Derived from a Doppler-free laser excitation study \citep{Niu2013}.}
\tablenotetext{b}{Derived from a VUV-FT study \citep{Niu2013}.}
\tablenotetext{c}{Derived from the semi-empirical model developed by \cite{Niu2013}.}
\tablenotetext{d}{Derived from lifetimes measured by \cite{Vansprang1977}.}
\end{deluxetable}

\subsection{The B-X system} \label{subsec:bcsyst}
\cite{Drabbels1993b} determined the rest wavelengths of the \bx\ band with an accuracy of \mbox{$0.003$ cm$^{-1}$}. The band oscillator strength $f_{\vv}$ used to calculate the oscillator strength values according to \mbox{Eq. (\ref{eq_fjj})} were obtained by averaging the values of \cite{Chan1993} and \cite{Zhong1998}, both measured via electron scattering, and \cite{Federman2001} and \cite{Stark2014}, who measured the $f_{\vv}$ values using optical absorption experiments. The sensitivity coefficients were calculated according to \mbox{Eq. (\ref{ki_pure})}, and the excited state lifetime \mbox{$\tau = 29.3 \pm 1.6$ ns}, used to calculate the damping parameter using \mbox{Eq. (\ref{eq_gamma_def})}, was measured by \cite{Drabbels1993a}. This value was found to be in good agreement with data from \cite{Krishnakumar1986}. The molecular parameters for the \bx\ band are listed in \mbox{Table \ref{bx_band}}. 

\begin{deluxetable}{cllccrcc}
\tablecaption{Molecular parameters for the B-X$ (0-0)$ band. \label{bx_band}}
\tablewidth{0pt}
\tablehead{
\colhead{\emph{J''}} & \multicolumn{2}{c}{Wavelength [\AA]} & \multicolumn{2}{c}{f$_{\jj}$} & \multicolumn{2}{c}{\emph{K$_{i}$}} & $\gamma_{\nuprime}$\tablenotemark{a} [s$^{-1}$] \\
 & \colhead{R} & \colhead{P} & \colhead{R} & \colhead{P} & \colhead{R} & \colhead{P} &
}
\startdata
0 & 1150.48254(13) & \multicolumn{1}{c}{-} & 6.62E-03 & - & $-0.00012$ & - &  0.34$\times 10^{8}$ \\
1 & 1150.43039(3) & 1150.58513(3) & 4.41E-03 & 2.21E-03 & $-0.00008$  & $-0.00021$ & \\
2 & 1150.37746(3) & 1150.63531(13) & 3.97E-03 & 2.65E-03 & $-0.00003$  & $-0.00026$ & \\
3 & 1150.32399(3) & 1150.68482(3) & 3.78E-03 & 2.84E-03 & 0.00001 & $-0.00030$ & \\
4 & 1150.26961(3) & 1150.73368(3) & 3.68E-03 & 2.94E-03 & 0.00006 & $-0.00034$ & \\
5 & 1150.21457(3) & 1150.78202(3) & 3.61E-03 & 3.01E-03 & 0.00011 & $-0.00038$ & \\
\enddata
\tablecomments{The uncertainties on the sensitivity coefficients are estimated to be better than $\sim 1\%$.}
\tablenotetext{a}{Derived from lifetimes measured by \cite{Drabbels1993a}.}
\end{deluxetable}

\subsection{The C-X system} \label{subsec:cxsyst}
The \cx\ band rest wavelengths adopted in this analysis were measured by \cite{Drabbels1993b} with an overall accuracy of \mbox{$0.003$ cm$^{-1}$}. The band oscillator strengths $f_{\vv}$ were measured by \cite{Federman2001} and \cite{Stark2014}, using optical absorption, and by \cite{Chan1993} and \cite{Zhong1998}, using electron scattering. The weighted average value from these studies was adopted in this work to calculate the $f_{\jj}$ values. The sensitivity coefficients were calculated according to \mbox{Eq. (\ref{ki_pure})}. \cite{Cacciani2001} used the time domain pump-probe technique to investigate the excited state lifetime of the C$^{1} \Sigma^{+}$, $\nuprime =0$ level and showed that it is not predissociated. They measured a value of \mbox{$\tau =  1.78 \pm 0.10$ ns} for the \cx\ band, which was converted into a value for the damping parameter $\gamma_{\nuprime J\,'}$. The molecular parameters for the \cx\ band are listed in \mbox{Table \ref{cx_band}}. 

\begin{deluxetable}{cllccccc}
\tablecaption{Molecular parameters for the \cx\ band.\label{cx_band}}
\tablewidth{0pt}
\tablehead{
\colhead{\emph{J"}} & \multicolumn{2}{c}{Wavelength [\AA]} & \multicolumn{2}{c}{f$_{\jj}$} & \multicolumn{2}{c}{\emph{K$_{i}$}} & $\gamma_{\nuprime}$\tablenotemark{a} [s$^{-1}$] \\
 & \colhead{R} & \colhead{P} & \colhead{R} & \colhead{P} & 
 }
\startdata
0 & 1087.86761(4) & \multicolumn{1}{c}{-} & 1.14E-01 & - & 0.000055 & - & 5.6$\times 10^{8}$ \\
1 & 1087.82110(4) & 1087.95910(4) & 7.61E-02 & 3.80E-02 & 0.000097 & $-0.000029$ & \\
2 & 1087.77413(4) & 1088.00408(4) & 6.85E-02 & 4.57E-02 & 0.000140 & $-0.000070$ & \\
3 & 1087.72668(4) & 1088.04871(4) & 6.52E-02 & 4.89E-02 & 0.000184 & $-0.000111$ & \\
4 & 1087.67876(4) & 1088.09263(4) & 6.34E-02 & 5.07E-02 & 0.000228 & $-0.000152$ & \\
5 & 1087.63014(10)& 1088.13620(10)& 6.23E-02 & 5.19E-02 & 0.000272 & $-0.000192$ & \\
\enddata
\tablecomments{The uncertainties on the sensitivity coefficients are estimated to be better than $\sim 1\%$.}
\tablenotetext{a}{Derived from lifetimes measured by \cite{Cacciani2001}.}
\end{deluxetable}

\subsection{The E-X system} \label{subsec:ecsyst}
The rest wavelengths for the E-X (0,0) band, as listed by \cite{Morton1994}, were in part adopted in this work for the R and P-branches. Much more accurate values for wavelengths of Q-branch lines were later measured by \cite{Cacciani2004}. The oscillator strengths were calculated starting from the weighted average of the $f_{\vv}$ values reported by \cite{Chan1993}, \cite{Zhong1998}, \cite{Federman2001}, and \cite{Stark2014}, while the sensitivity coefficients were calculated using \mbox{Eq. (\ref{ki_pure})}. As for the natural lifetime damping coefficients it should be considered that the E$^1\Pi, v=0$ state is predissociated \citep{Cacciani1995,Cacciani1998}, and that both radiative and predissociative decay contribute to the natural lifetime broadening, resulting in a value of \mbox{$\tau = 0.91 \pm 0.06$ ns} for the excited state lifetime. This value was converted into a value for $\gamma_{\nuprime J\,'}$ using \mbox{Eq.(\ref{eq_gamma_def})}. The molecular parameters for the \ex\ band are listed in \mbox{Table \ref{ex_band}}.
\begin{deluxetable}{clllcccrccc}
\rotate
\tablecaption{Molecular parameters for the \ex\ band. \label{ex_band}}
\tablewidth{0pt}
\tablehead{
\colhead{\emph{J"}} & \multicolumn{3}{c}{Wavelength [\AA]} & \multicolumn{3}{c}{f$_{\jj}$} & \multicolumn{3}{c}{\emph{K$_{i}$}} & $\gamma_{\nuprime}$\tablenotemark{a} [s$^{-1}$] \\
 & \colhead{R} & \colhead{Q} & \colhead{P} & \colhead{R} & \colhead{Q} & \colhead{P} & \colhead{R} & \colhead{Q} & \colhead{P} & 
 }
\startdata
0 & 1076.03361(12) & - & - & 6.44E-02 & - & - & $-0.00010$ & - & - & 10.96$\times 10^{8}$ \\
1 & 1075.98718(12) & 1076.07891(3) & - & 3.22E-02 & 3.22E-02 & - & $-0.00006$ & $-0.00014$ & - & \\
2 & 1075.93960(12) & 1076.07751(3) & 1076.16713(12) & 2.58E-02 & 3.220E-02 & 6.44E-03 & $-0.00001$ & $-0.00014$ & $-0.00022$ & \\
3 & 1075.89133(12) & 1076.07540(3) & 1076.20975(12) & 2.30E-02 & 3.220E-02 & 9.20E-03 & 0.00003 & $-0.00014$ & $-0.00026$ & \\
4 & 1075.84202(12) & 1076.07261(3) & 1076.25133(12) & 2.15E-02 & 3.220E-02 & 1.07E-02 & 0.00008 & $-0.00013$ & $-0.00030$ & \\
5 & 1075.79167(12) & 1076.06913(3) & 1076.29199(12) & 2.05E-02 & 3.220E-02 & 1.17E-02 & 0.00012 & $-0.00013$ & $-0.00034$ & \\
\enddata
\tablecomments{The uncertainties on the oscillator strengths are estimated to be better than \mbox{$\sim 10\%$}, and the sensitivity coefficients are estimated to be better than $\sim 1\%$.}
\tablenotetext{a}{Derived from lifetimes measured by \cite{Cacciani1998}.}
\end{deluxetable}

\section{Quasar absorption model} \label{sec:model}
Quasar J1237+0647 is located at the emission redshift \mbox{$\zem = 2.78$} and in its line-of-sight is located a sub-damped Lyman-$\alpha$ system, \mbox{$\log[N_{\textrm{\scriptsize H\,I}}/{\rm cm}^{-2}] = 20.00 \pm 0.15$}
, featuring atomic and molecular absorption features at an absorption redshift of \mbox{$\zabs = 2.69$} \citep{Noterdaeme2010}. Molecular hydrogen and deuterated molecular hydrogen (HD) absorption features were investigated by \cite{Dapra2015}, who used H$_{2}$ and HD to derive a constraint on variation in $\mu$. Here is presented the analysis of the CO absorption in the spectrum of J1237+0647 and the constraint on $\mu$ that it delivers.

\subsection{The fitting method} \label{subsec:abs_model}
The absorption model was created using the non-linear least-squares Voigt profile fitting program \textsc{vpfit} \citep{vpfit}. Within the program, a Voigt profile is described by a set of three free parameters, which are used together with the molecular parameters: the column density \emph{N}, the redshift at which the absorption occurs \emph{z}, and the Doppler line width \emph{b}. A comprehensive fit was performed, involving a simultaneous treatment of all the lines \citep{Malec2010,King2011,Bagdonaite2014,Dapra2015}. The main strength of this method is that each free parameter can be shared by different transitions, thus minimizing the number of free parameters needed to perform the fit. This allows one to fit even the molecular absorption features that show partial overlaps with intervening lines from metals and \ion{H}{1}, as well as to deal with the blending of the P, Q and \mbox{R branches} of the CO features that the weak transitions involving high rotational levels, up to $J = 5$.

\begin{figure}
\centering
\includegraphics[width=\columnwidth]{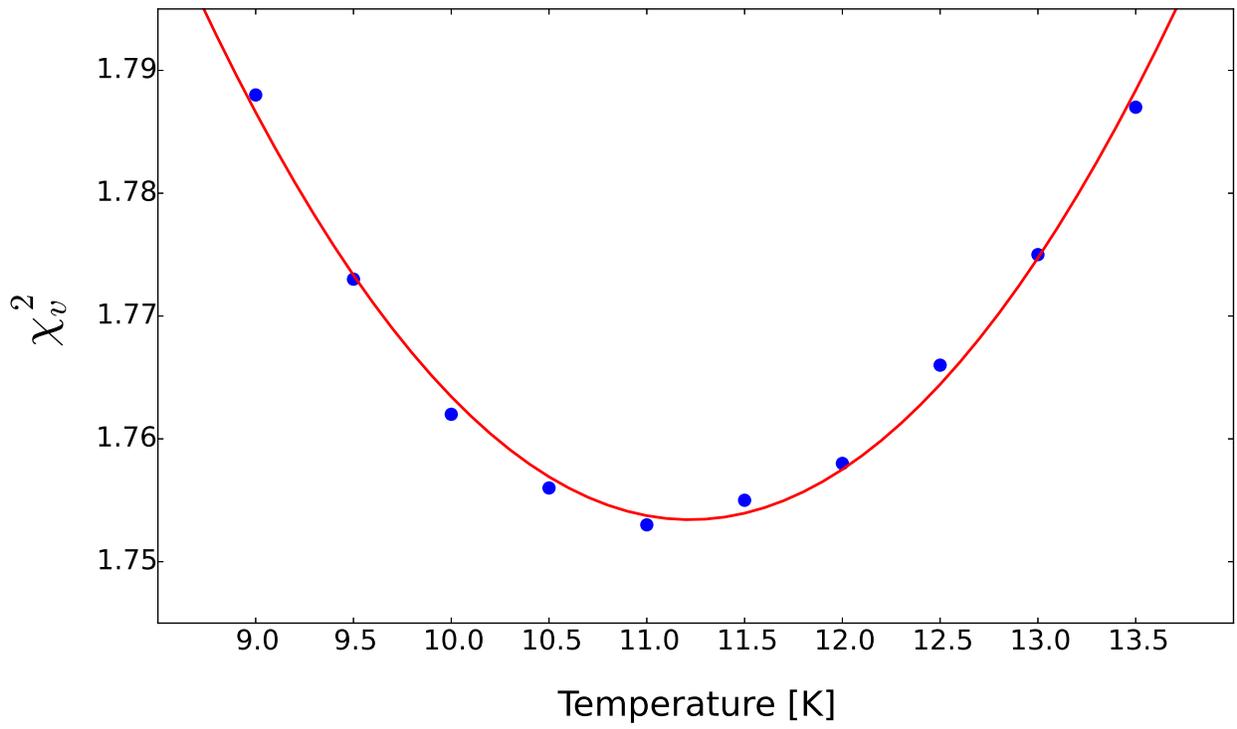}
\caption{Reduced $\chi^{2}$ from fitted CO models with different temperatures. The values of the $\chi^{2}_{\nu}$ are indicated with (blue) dots and the best fit is presented with a (red) solid line. \label{fig_co_t}}
\end{figure}

Since the CO transitions originate in the same absorber and they are assumed to share the same physical conditions of the absorbing cloud, the redshift \emph{z} and the width \emph{b} parameters were tied together. The rotational state-dependent column densities $N_{J}$ were linked together assuming thermodynamic equilibrium yielding a Boltzmann distribution at a temperature \emph{T}:
\begin{equation}
N_{J} = N_{col} P_{J}(T) = N_{col} \frac{(2J + 1) e^{-E_{rot}/kT}}{\sum{(2J + 1)}},
\label{n_j}
\end{equation}
where $N_{col} = \sum{N_{J}}$ is the total column density for CO, and $P_{J}(T)$ is the partition function giving the relative population of the single rotational \emph{J}-states. The gas temperature $T_{\textrm{\scriptsize{CO}}}$ was used to calculate the partition function, but was not treated as a free parameter in \textsc{vpfit}. Models corresponding to different CO temperatures were fitted to the spectrum in multiple runs, under the assumption that CO is in thermodynamic equilibrium, resulting in the best-fit temperature \mbox{$T_{\textrm{\scriptsize{CO}}} = 11.2 \pm 0.1$ K}, as shown in \mbox{Fig. \ref{fig_co_t}}. This temperature is close to the expected temperature due to the excitation of the cosmic microwave background (CMB) \mbox{$T_{\textrm{\scriptsize{CMB}}}(\zabs = 2.69) = 10.05$ K} and verifies the assumption of equilibrium for the population distribution \citep{Noterdaeme2010}. The procedure results in a model which consists of only three free parameters describing the CO transitions: the redshift \emph{z}, the line width \emph{b}, and the total column density of the gas \emph{N$_{col}$}. Effectively, a shared vibrational band contour of overlapping CO lines is fitted, rather than multiple, individual rovibronic line profiles.

It is noted here that while CO and H$_{2}$ are the main molecular constituents of interstellar clouds in galactic media, their behaviour and thermodynamic properties are usually very different. The CO gas, as observed in the high-redshift absorbing systems, is found to exhibit thermalized population distributions at the local cosmic microwave background temperature \citep{Noterdaeme2010}. In contrast, the population distribution of H$_{2}$ molecules is non-thermal with higher rotational states populated superthermal. For the lowest levels a temperature-like distribution is found with $T_{01} \sim 50-100$ K for the lowest two rotational levels \citep{Petitjean2002}. Also the observed widths of the absorption lines in all high-redshift extra-galactic objects exceeds the kinetic temperatures that would correlate with the Boltzmann temperatures. These widths, treated with a Doppler parameter \emph{b}  in studies probing varying constants, are ascribed to turbulent motions in the observed clouds. For this reason the physical parameters \emph{b} do not represent a temperature, nor can be they be equated for the different species observed.

\subsection{CO bands} \label{subsec:co_bands}
The CO bands pertaining to the A-X system fall in the red part of the spectrum, at redshifted wavelengths \mbox{$\lambda > 4877$ \AA}, and their absorption profiles do not show significant overlaps with any other spectral feature. Nine spectral regions were selected in the range \mbox{$\lambda = 4877 - 5702$ \AA} covering the A-X bands. The region containing the A-X$(1 - 0)$ band also includes the perturbing d-X$(5 - 0)$ band, which however is not overlapping the A-X$(1 - 0)$ band. Some metal absorption features, namely \ion{Si}{4} at $\zabs \simeq 2.69, 2.62$ and \ion{C}{4} at $\zabs \simeq 2.59$, fall near the CO bands, in which cases the atomic lines were included in the fit in order to obtain a better constraint on the continuum level close to the CO features. From the Franck-Condon factor analysis, the A-X$(9 - 0)$ band is strong enough to be detected, but it is almost completely overlapped by strong metal features at \mbox{$\lambda \sim 4802$ \AA} and it was not included in the model.

The other CO absorption bands in the spectrum of J1237+0647, C-X, B-X, and E-X, are detected in the blue arm of UVES at redshifted wavelengths shorter than the Lyman-$\alpha$ emission feature of the quasar. This region is referred to as the Lyman-$\alpha$ forest and shows multiple \ion{H}{1} absorption features arising from the intergalactic medium at redshifts \mbox{$z < \zem$}. It is common that such neutral hydrogen lines overlap absorption features falling in the Lyman-$\alpha$ forest. In such cases, the overlapping \ion{H}{1} features occurring in the selected CO regions were included in the model by assigning to each of them a set of free parameters in \textsc{vpfit}.

The B-X electronic system falls in the Lyman-$\alpha$ forest at redshifted wavelengths \mbox{$\lambda < 4246$ \AA}. Of its three known vibrational levels, only the \bx\ is strong enough to be detected in this absorber in the region \mbox{$\lambda = 4242 - 4246$ \AA}. Its \mbox{R branch} is partially overlapped by an intervening saturated \ion{H}{1} line at \mbox{$\lambda \sim 4243.7$ \AA}, and an additional narrow, unidentified feature at \mbox{$\lambda \sim 4243.1$ \AA}.

Of the four known vibrational levels of the C-X electronic system, only the \cx\ band was detected in the region \mbox{$4243 - 4246$ \AA}. It is partially overlapped, mainly in its \mbox{R branch}, by a saturated \ion{H}{1} line occurring at \mbox{$\lambda \sim 4013$ \AA}. The C-X$(1 - 0)$ band has an oscillator strength which is of the same order as the \bx\ band; however, the former falls in a heavily saturated region, thus is not detected.The C-X$(2- 0)$ and $(3 - 0)$ bands have an oscillator strength \mbox{$\sim 10^{4}$} times weaker than the \cx\ band \citep{Morton1994}, hence are not detected. 

The E-X electronic system is detected in the J1237+0647 spectrum at wavelengths \mbox{$\lambda < 3972$ \AA}. Compared to \ex, the E-X$(1-0)$ band is one order of magnitude weaker, while the E-X$(2-0)$ band is two orders of magnitude weaker \citep{Morton1994}, hence these bands are too weak to be detected. The transitions of the Q branch of the \ex\ band are overlapped, resulting in a clear feature at \mbox{$\sim 3970.3$ \AA}, while an intervening \ion{H}{1} line partially overlaps the \mbox{P branch} at $\lambda \simeq 3971$ \AA. 

CO has two more electronic systems, V-X at a rest wavelength \mbox{$\lambda \sim 1011$ \AA}, and F-X at \mbox{$\lambda \sim 1003$ \AA}. While the V-X$(0-0)$ band is too weak to be detected, the F-X$(0-0)$ band is stronger \citep[\mbox{$\sim 2$ times} stronger than the \bx\ band,][]{Morton1994}. However, the F-X$(0 - 0)$ is completely overlapped by strong \ion{H}{1} lines and cannot be identified in this absorber. CO has many more absorption systems lying at \mbox{$\lambda < 1000$ \AA} that are generally weaker and not included in the present discussion \citep{Eidelsberg1990,Eikema1994}.

\begin{figure}
\centering
\includegraphics[scale=0.3]{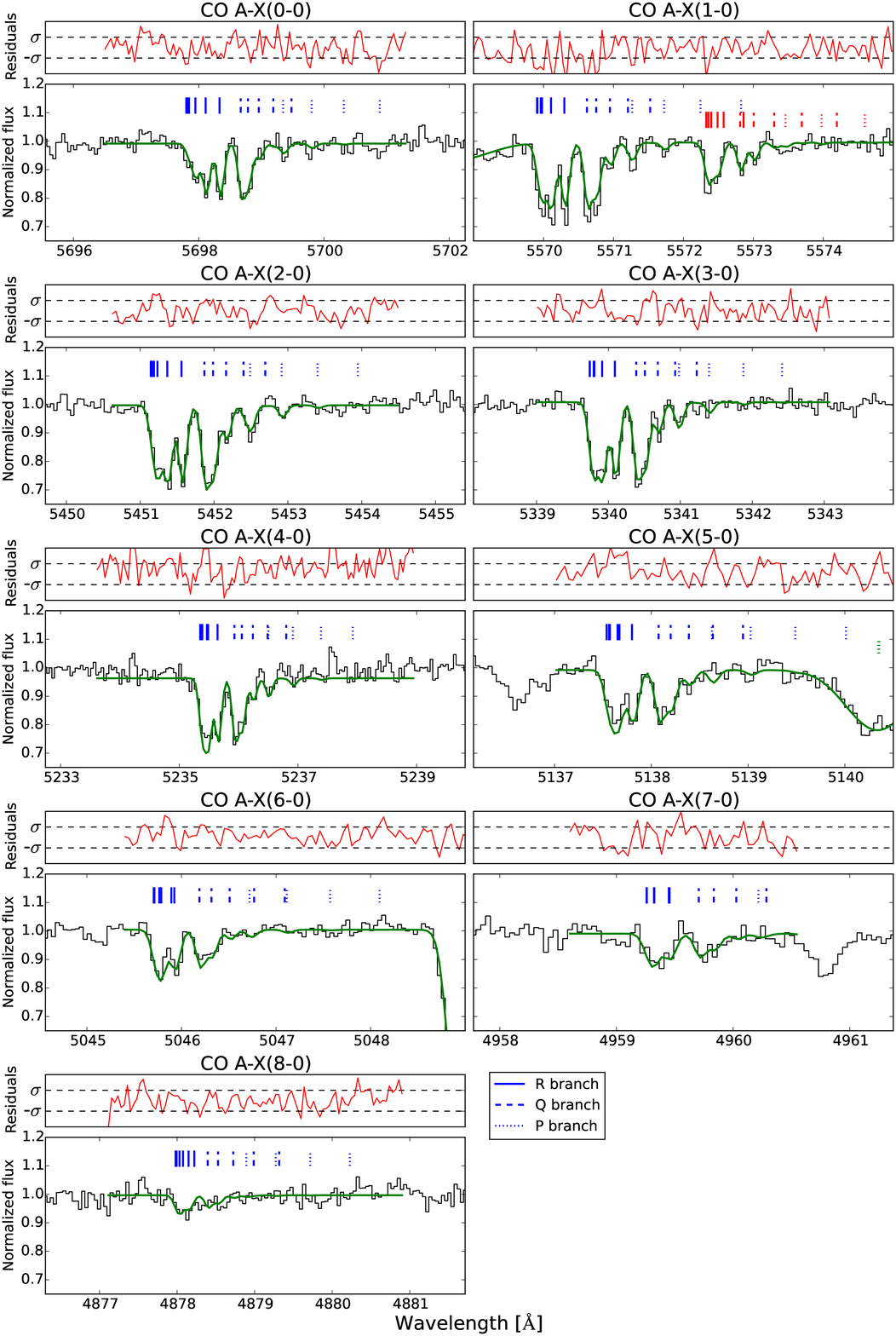}
\caption{Absorption model for the CO bands from A-X$(0-0)$ to A-X$(8 - 0)$. The (green) solid line represents the fitted model while the (blue) ticks show the wavelengths of the rotational lines for ground states \mbox{$J = 0-5$} and their different branches. Band A-X$(1-0)$ is perturbed by the inter-system band d-X$(5-0)$, whose rotational levels are shown by the (red) ticks. Band A-X$(5-0)$ falls close to a \ion{Si}{4} absorption feature at \mbox{$\lambda \sim 5140.5$ \AA}, which is indicated by the (green) dotted tick. The (red) solid lines represent the residuals of the fits with their $\pm 1 \sigma$ boundaries. \label{ax_bands}}
\end{figure}

\begin{figure}
\centering
\includegraphics[scale=0.3]{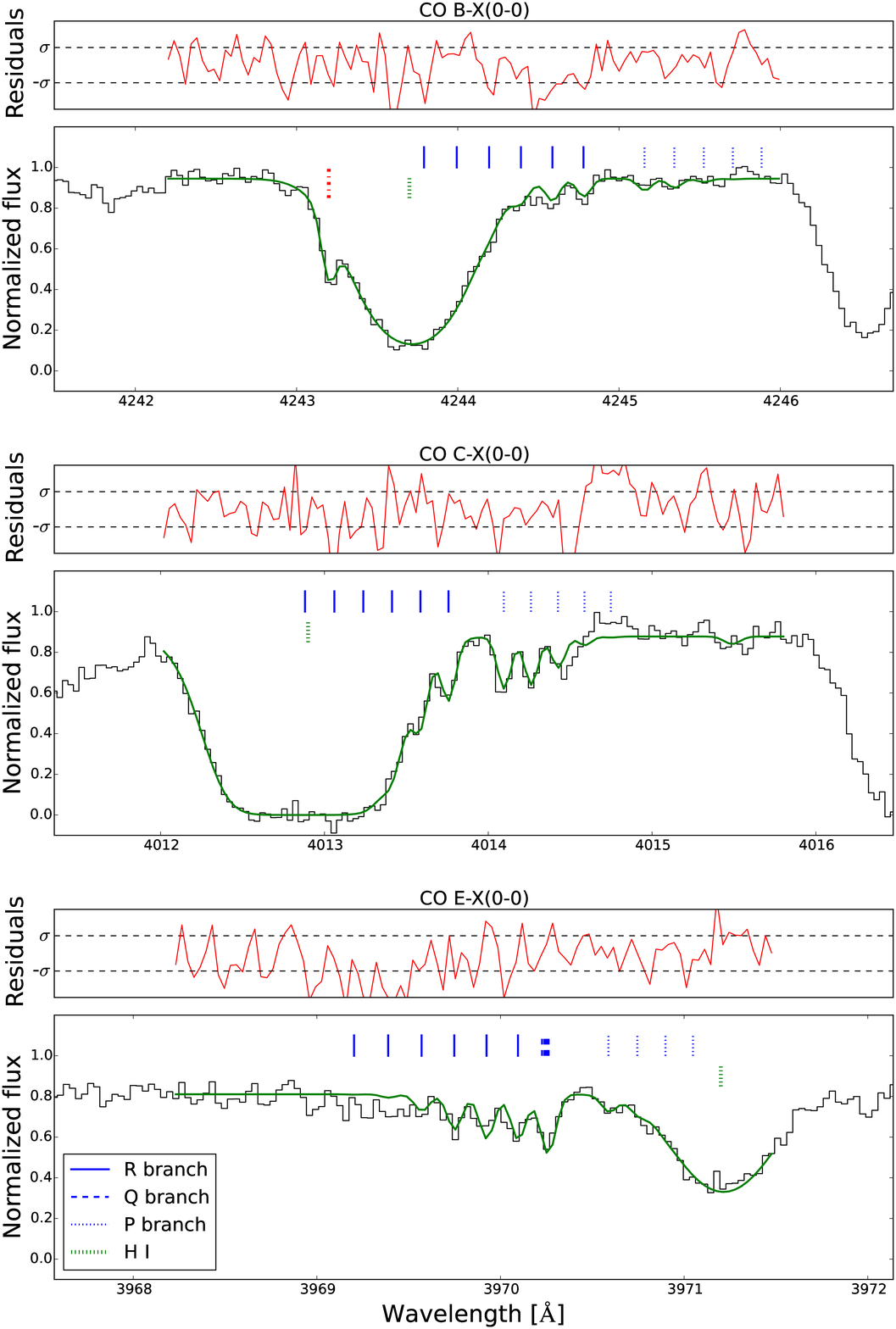}
\caption{Top panel: absorption model for the \bx\ band. Middle panel: absorption model for the \cx\ band. Bottom panel: absorption model for the \ex\ band. The (green) solid line represents the fitted model while the (blue) ticks show the wavelengths of the rotational lines for ground states \mbox{$J = 0-5$} and their different branches. The (red) solid lines represent the residuals of the fits with their $\pm 1 \sigma$ boundaries. The (green) dashed ticks show the positions of the intervening \ion{H}{1} lines. The (red) dashed tick at $\lambda \sim 4243.1$ shows the position of an unidentified narrow absorption feature. \label{co_blue_bands}}
\end{figure}

For CO absorption, a total column density of \mbox{$\log [N/\textrm{cm}^{-2}] = 14.29 \pm 0.02$}, an absorption redshift of \mbox{$\zabs = 2.689566(1)$} and a width of \mbox{$b = 0.73 \pm 0.03$ \kms} are obtained from the fit. These values agree with those of \cite{Noterdaeme2010} within \mbox{$1.5 \sigma$} significance. Moreover, the normalized residuals are distributed in the range \mbox{$\pm 1\sigma$} in each fitted spectral region, validating the assumption that there is no overlap with other undetected spectral features. The absorption model for the CO bands is presented in \mbox{Fig. \ref{ax_bands}} for the A-X system, and in \mbox{Fig. \ref{co_blue_bands}} for the \bx, \cx, and \ex\ bands.

The reduced chi-squared parameter returned by the model is \mbox{$\chi^{2}_{\nu} = 1.5$} with \mbox{$\nu = 7391$} degrees of freedom. The main contributors to the final $\chi^{2}_{\nu}$ value being somewhat larger than unity are the CO bands in the blue arm; indeed the fit of only the A-X bands returns a \mbox{$\chi^{2}_{\nu} = 1.22$}. This is due to the fact that the \bx, \cx, and \ex\ bands show overlaps with intervening, saturated \ion{H}{1} lines at \mbox{$\lambda \sim 4243.5$} and \mbox{$4013$ \AA}. Another contributor is the region that includes the A-X$(1 - 0)$ and the d-X$(5 - 0)$ bands. Both these bands appear slightly stronger in the spectrum compared to the absorption model, while the other A-X$(\nuprime - 0)$ bands appear slightly weaker than predicted by the model. This may be related to possible, small errors in the values of their band oscillator strengths included in the CO molecular database. Excluding the spectral regions containing these CO bands from the absorption model delivers a reduced chi-squared of \mbox{$\chi^{2}_{\nu} = 1.2$}.

\begin{figure}
\centering
\includegraphics[width=\columnwidth]{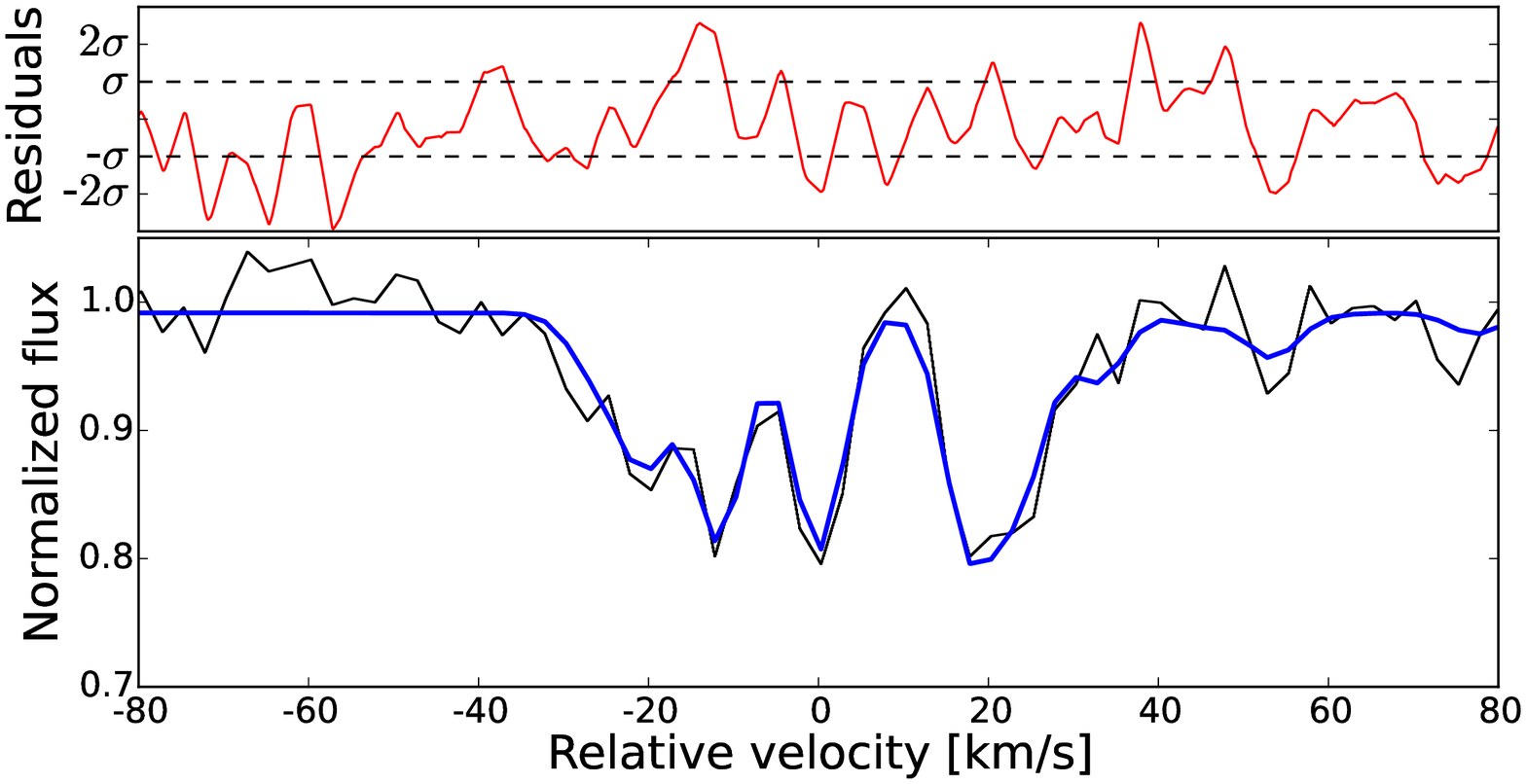}
\caption{Top panel: normalized composite residual spectrum from the 13 CO bands detected. Bottom panel: the absorption model for the CO A-X$(1- 0)$ band is plotted as a reference. The (blue) solid line represents the fitted model. The velocity scale is centred on the R(0) transition for both the panels. \label{crs}}
\end{figure}

Another reason for a \mbox{$\chi^{2}_{\nu} > 1$} may be the possible presence of extra velocity components that were not included in the model. To investigate this, a composite residual spectrum \citep[CRS,][]{Malec2010} was created by combining the residual structure between the spectrum and the model for all the CO bands. The CRS, as shown in \mbox{Fig. \ref{crs}}, does not show evidence for missing velocity components in the CO absorption model. Moreover, multiple trial models with two CO velocity components were fitted to the spectrum, but the second, weaker velocity component was rejected by \textsc{vpfit}. As a consequence, the presence of an extra CO velocity component was excluded. 

Finally, small shifts in individual transitions are likely to cause discrepancies between the recorded spectrum and the absorption model, and will be another reason for a final $\chi^{2}_{\nu} > 1$. Such shifts are caused by wavelength calibration distortions on scales of single echelle orders. The effect of these intra-order distortions is discussed in \mbox{Section \ref{subsubsec:intra-order}} and is included in the systematic error budget. The uncertainties on the CO fitting parameters $\log N$, \emph{z}, and \emph{b} were scaled by the square root of the final $\chi^{2}_{\nu}$ value, in order to take into account the discrepancies between the model and the spectrum. The aforementioned phenomena originate as well the larger $\chi^{2}_{\nu}$ value of the fit to determine the CO population temperature (see \mbox{Fig. \ref{fig_co_t}}).

\subsection{Constraining $\dmm$} \label{subsec:dmm}
A variation of the proton-to-electron mass ratio can be probed using the absorption spectra of rovibronic molecular transitions. A change in the value of $\mu$ will be reflected in a shift of the observed wavelengths. This shift, which is assumed to be linear, is given by:
\begin{equation}
\lambda^{obs}_{i} = \lambda^{rest}_{i} (1 + \zabs)(1 + K_{i}\frac{\Delta\mu}{\mu}),
\label{eq_wl_shift}
\end{equation}
where $\lambda^{obs}_{i}$ is the observed wavelength of the \emph{i}-th transition, $\lambda^{rest}_{i}$ is its rest wavelength, and $\zabs$ is the redshift at which the absorption occurs. $\dmm = (\mu_{z}-\mu_{lab})/\mu_{lab}$ is the relative difference between the value of the proton-to-electron mass ratio in the absorption system, $\mu_{z}$, and the one measured on Earth, $\mu_{lab}$, and $K_{i}$ is the sensitivity coefficient specific for each transition \emph{i}.

\begin{figure}
\includegraphics[width=\columnwidth]{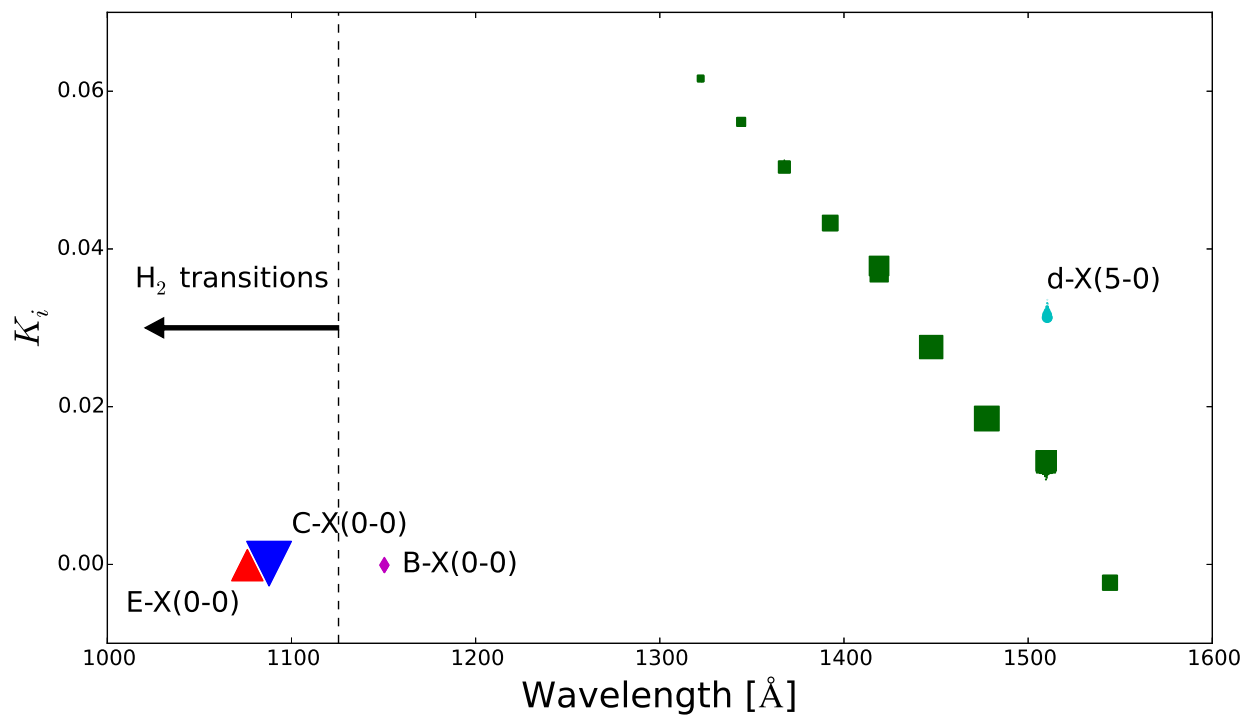}
\caption{Sensitivity coefficients of the CO bands detected in the blue arm (left panel) and in the red arm of UVES (right panel). A-X bands are shown with (green) squares, the \mbox{d-X$(5-0)$} band with (cyan) circles, the \bx\ with (magenta) diamonds, the \cx\ with (blue) downward-pointing triangles, and the \ex\ with (red) upward-pointing triangles. The size of each marker is proportional to the line intensity. Note that the `tear drop' shape of the \mbox{d-X$(5 - 0)$} and some A-X bands inserts is a consequence of contributions by a number of rotational lines in the bands, some of which undergo perturbations. The H$_{2}$ and HD transitions fall in the area shown by the (black) arrow, delimited by the dashed line. \label{fig_co_ki}}
\end{figure}

It follows from \mbox{Eq. (\ref{eq_wl_shift})} that, if the sensitivity coefficients are wavelength-dependent, the presence of a wavelength distortion could mimic a shift in $\mu$ \citep{Ivanchik2005,Ubachs2007,Malec2010}. The CO bands, whose $K_{i}$ coefficients are shown in \mbox{Fig. \ref{fig_co_ki}}, show a wavelength dependence only in the A-X electronic system, while the \bx, \cx, and \ex\ bands all have sensitivity coefficients \mbox{$\sim 0$} at rest wavelengths of \mbox{$\lambda \sim 1075 - 1151$ \AA} and were used as anchor transitions in this analysis. The degeneracy can be further broken by including in the model the d-X$(5 - 0)$ band, since it has very different coefficients than the A-X$(1 - 0)$ band at similar wavelength.

After having developed a robust absorption model, an extra free parameter was introduced, beside the set of parameters describing the CO absorption, in a final fitting run in \textsc{vpfit} in order to constrain the $\mu$-variation. The extra parameter $\dmm$ was not introduced earlier to avoid that a wrong estimate of the absorption redshift was compensated by an artificial $\mu$ variation caused by the degeneracy between the redshift and a non-zero $\dmm$. The model returned a constraint on the variation of the proton-to-electron mass ratio of \mbox{$\dmm = (0.7 \pm 1.6_{stat}) \times 10^{-5}$}, hereafter referred to as the fiducial value of $\dmm$. The statistical error is derived from the diagonal term of the final covariance matrix for the fit, and it represents only the uncertainty in $\dmm$ derived from the S/N of the quasar spectrum. The statistical error derived from CO is  larger than the error obtained from the previous analysis of H$_{2}$ absorption in the same system by a factor of \mbox{$\sim 3$} \citep{Dapra2015}.

\section{Systematic uncertainty} \label{sec:systematics}
An estimation of the systematic error affecting the constraint on $\dmm$ derived from CO absorption only was made considering the contributions to the error budget from the five most dominant sources and discussed extensively below.

\subsection{Wavelength scale distortions} \label{subsec:wav_distortions}
In recent years the UVES spectrograph was found to suffer from wavelength calibration distortions both on scales of single echelle orders \citep{Griest2010,Whitmore2010,Whitmore2015} and longer scales \citep{Rahmani2013,Whitmore2015}. These distortions are most likely due to different light paths between the object observed during the science exposures and the ThAr lamp located on the VLT platform \citep{Molaro2008}. The long-range distortions in particular would introduce a wavelength-dependent velocity shift which is nearly degenerate with a non-zero $\dmm$, as discussed in \mbox{Section \ref{subsec:dmm}}. In principle such degeneracy can be broken by fitting together CO bands that have different sensitivity coefficients at similar wavelengths, as in the case of the \mbox{A-X$(1- 0)$} and the \mbox{d-X$(5 - 0)$} bands, or bands whose $K_{i}$ are not wavelength dependent, like the \bx, \cx, and \ex\ bands. Excluding the \mbox{d-X$(5 - 0)$} and the three anchor bands from the fit results in a constraint of \mbox{$\dmm = (-0.4 \pm 2.1_{stat}) \times 10^{-5}$}, whose uncertainty is \mbox{$\sim 35\%$} larger than the fiducial value, indicating some effectiveness in breaking this degeneracy. However, the three anchor bands are overlapped by intervening \ion{H}{1} lines, reducing the effectiveness of this approach in breaking the degeneracy.

\subsubsection{Long-range distortions} \label{subsubsec:long_range}
To account for the calibration distortions, the technique now referred to as `supercalibration' was first demonstrated by \cite{Molaro2008} and later improved by \cite{Rahmani2013} and, more recently, \cite{Whitmore2015}, whose method was used to supercalibrate the spectrum of J1237+0647. The `supercalibration' technique consists in the comparison of a ThAr-calibrated UVES spectrum to a reference Fourier-transform absorption spectrum with a much more accurate frequency scale \citep{Chance2010}\footnote{Available at \url{http://kurucz.harvard.edu/sun/irradiance2005/irradthu.dat}}. Targets for supercalibrations are asteroids, which reflect the solar light and hence show the same spectrum of the Sun, and `solar twin' stars, which are objects with a spectrum that is almost identical to the solar one \citep{Melendez2009,Datson2014}. The spectrum of J1237+0647 was partially corrected for long-range distortions following the same supercalibration procedure used by \cite{Dapra2015}.

\begin{figure}
\centering
\includegraphics[scale=0.45]{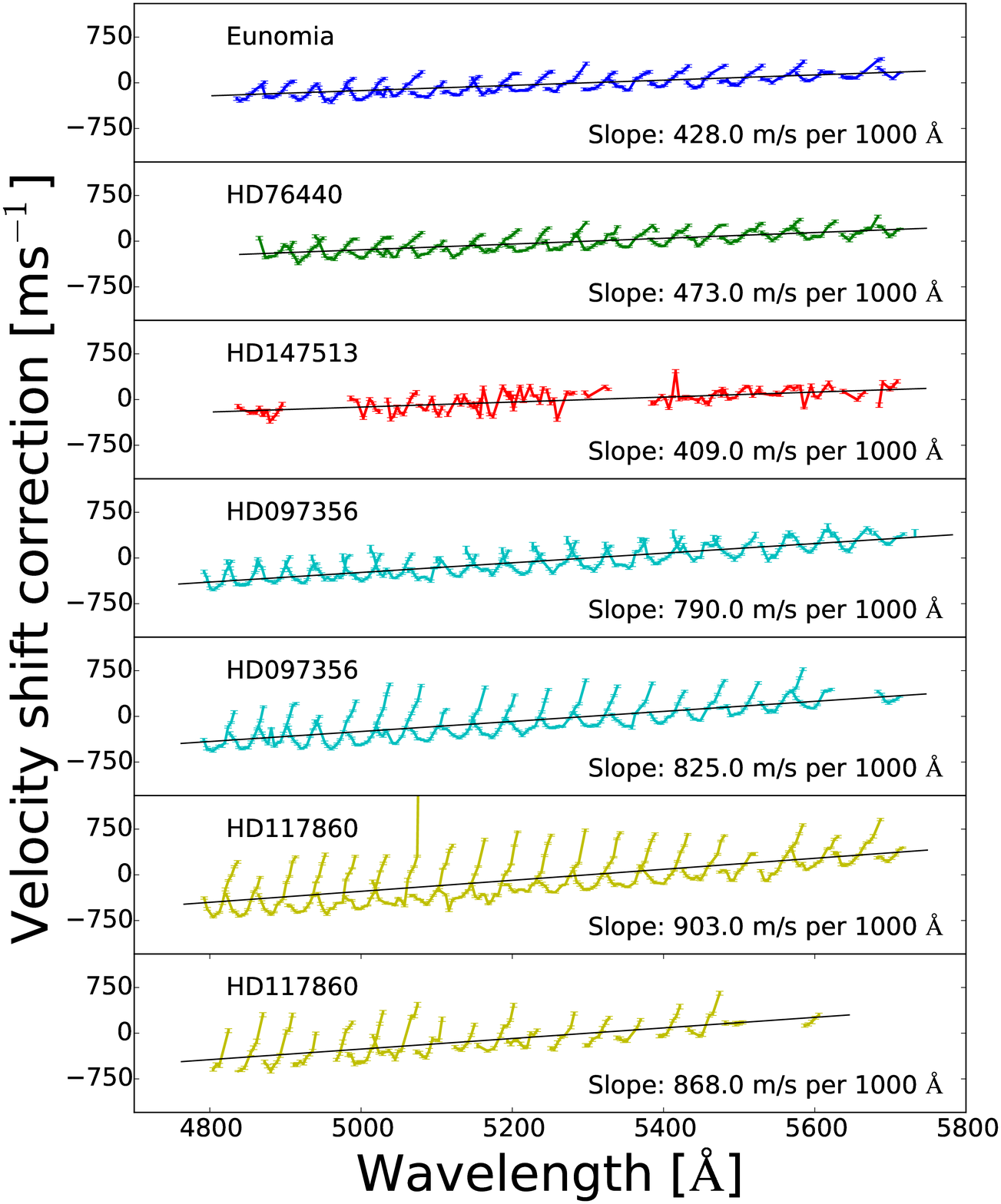}
\caption{Distortions in exposures of one asteroid (Eunomia) and six `solar twin' stars, of the UVES wavelength scale used to supercalibrate the exposures of quasar J1237+0647 taken in 2013 and 2014. For each exposure, the velocity shift measurements are made on \mbox{$\sim 10$} echelle orders. The slopes of the fitted lines show the velocity shift needed to counter the effect of the long-range distortions. The values of the slopes are indicated for each supercalibration. The `solar twin' stars HD097356 and HD117860 were observed in service mode throughout the period March - May 2014, while the other targets were observed in visitor mode in May 2013. In all cases the supercalibrations exposures were taken immediately after the J1237+0647 science exposures. Since only the slopes are the physically relevant parameters for the supercalibration process, the distortions were shifted to a zero velocity shift at \mbox{$\lambda = 5300$ \AA}. \label{supercali_maps}}
\end{figure}

Exposures taken in 2013 and 2014, \mbox{for $\sim 11.5$ hrs} of integration, have dedicated supercalibrations for both the blue and the red arm of UVES. The supercalibrations for the spectrum of J1237+0647 in the red arm are presented in Fig. \ref{supercali_maps}, while the blue arm was calibrated as in \cite{Dapra2015}. The impact of the long-range distortions affecting the exposures taken in 2009, covering \mbox{$\sim 8.5$ hrs} of integration, was estimated using observations of the Ceres asteroid performed within one week of the quasar exposures (program 080.C-0881(B), PI Dumas). The Ceres exposures were taken using only the blue arm of UVES and yielded two distortion slopes of \mbox{$\sim 150$ \ms} per 1000 \AA\ and one of \mbox{$\sim -500$ \ms} per 1000 \AA. The positive slope value was used to correct for the distortions in the exposures taken in 2009 in the blue arm, while the negative value was translated into a systematic uncertainty on $\dmm$ of \mbox{$\sim 2.4 \times 10^{-6}$}, as in \cite{Dapra2015}. On average, the long-range distortions in the red arm of UVES have a slightly larger magnitude than what is measured in the blue arm. A comparison between the red and the blue arm returns, for the supercalibrations exposures taken in 2013 and 2014, returns an average ratio of $\sim 1.2$, with only the `solar twin' HD117860 delivering a larger ratio of $\sim 2.5$. The average ratio was used to estimate the magnitude of the long-range distortions in the red arm for the exposures taken in 2009. Another constraint of \mbox{$\dmm = (0.5 \pm 1.6_{stat}) \times 10^{-5}$} was derived using the larger ratio from HD117860, and a spread in $\dmm$ of \mbox{$\sim 2 \times 10^{-6}$} between the two constraints was added to the systematic error budget.

The `solar twins' HD097356 and HD117860 were observed in 2014 two times in a time window of 10 and 3 days respectively. These target show a variation of their distortions slopes of $+5\%$ and $-5\%$ respectively. The distortion corrections of each J1237+0647 exposure were first enhanced and subsequently decreased by $5\%$ to simulate the effect of a temporal variation of \mbox{$\sim 1$ week}. Two different constraints of \mbox{$\dmm = (0.6 \pm 1.6_{stat}) \times 10^{-5}$} and \mbox{$\dmm = (0.7 \pm 1.6_{stat}) \times 10^{-5}$} were derived from the positive and from the negative variation respectively. These constraints show that the temporal difference between the quasar exposures and the Ceres supercalibration exposures introduces an error of, at most, \mbox{$1 \times 10^{-6}$} on the fiducial value of the constraint.

\subsubsection{Intra-order distortions} \label{subsubsec:intra-order}
The presence of intra-order wavelength distortions introduces in each exposure a velocity shift which translates into a $\dmm$ uncertainty given by \mbox{$\delta(\dmm) = [(\Delta v/c)/\sqrt{N}]/\Delta K_{i}$}. Here, $\Delta v$ is the mean amplitude of the intra-order distortions, \mbox{$\Delta K_{i} = 0.06$} is the spread in the CO sensitivity coefficients, and \emph{N} is the number of CO transitions considered in the analysis. The mean amplitude of the intra-order distortions in the red arm is \mbox{$\Delta v = 520$ \ms} for exposures taken in 2013 and 2014. This amplitude is very similar to what was measured in the blue arm for all the exposures used in the analysis. Therefore, it was taken as the mean amplitude of the intra-order distortions for all exposures, including the ones taken in 2009.

Since the CO bands have mostly blended R and Q branches, while the P branches are weak, a band contour, including the higher \emph{J} levels was effectively fitted. The number of transitions containing valuable signal is reflected by the enhancement of the statistical precision on $\dmm$ after combining the constraint obtained from CO only with the constraint from 137 H$_{2}$ and HD transitions \mbox{$\dmm(\textrm{H}_{2}) = (-5.4 \pm 6.3_{stat} \pm 4.0_{syst}) \times 10^{-6}$}. A combined constraint (see \mbox{Section \ref{sec:combined}}) results in a statistical uncertainty on $\dmm$ which is \mbox{$\sim 10\%$} smaller, since the statistical error scales with $\sqrt N$, a value of \mbox{$N=36$} of CO transitions that are effectively contributing to the signal was adopted, resulting in an uncertainty on $\dmm$ of \mbox{$\sim 4.5 \times 10^{-6}$}.

\subsection{Uncertainty from using different UVES arms} \label{subsec:offset}
Another potential cause of error is the presence of an offset between the wavelength scales of the blue arm of UVES where the \bx, \cx, and \ex\ bands are covered, and the lower red arm where the A-X bands fall. To quantify this effect, metal absorption features detected both in the blue and in the red arms, were investigated. Since the considered transitions belong to the same atom, the redshift $\zabs$ at which they originate is expected to be the same for features detected in the two arms of UVES. Any non-zero offset $\Delta \zabs = \zabs^{\textrm{\scriptsize blue}} - \zabs^{\textrm{\scriptsize red}}$ is evidence that there is a shift between the wavelength scales of the two arms, and such shift would introduce an effect mimicking $\dmm \not= 0$.

\begin{figure}
\centering
\includegraphics[scale=0.65]{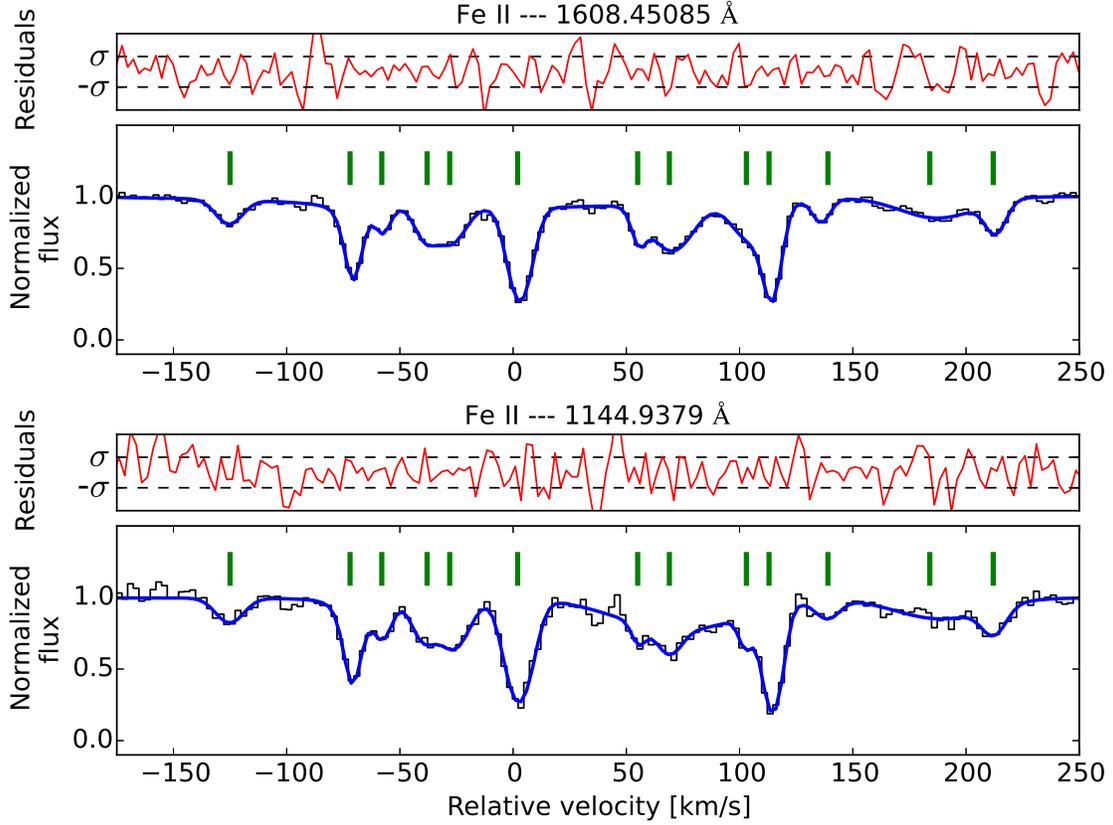}
\caption{Velocity profiles of two transitions of \ion{Fe}{2} in the red part (top panel) and in the blue part of the spectrum (bottom panel). The solid blue line shows the absorption models, and the green ticks show the position of the VCs. On top of each spectrum, residuals are shown with a red dashed line, and the two horizontal dashed lines show their $\pm\,1\sigma$ boundaries. The velocity scale is centred at redshift $z = 2.689570$. \label{fe2}}
\end{figure}

\begin{figure}
\centering
\includegraphics[scale=0.35]{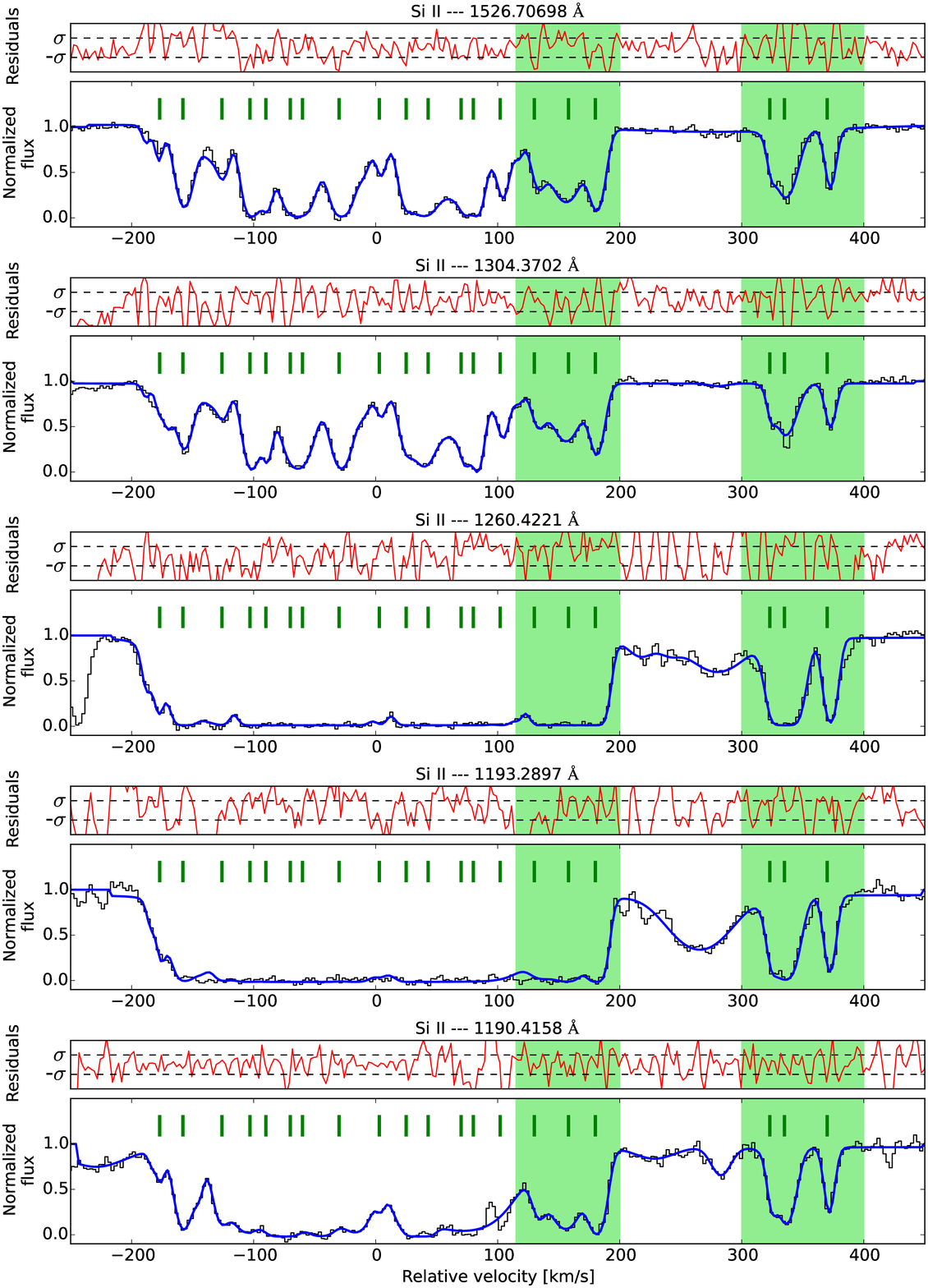}
\caption{Velocity profiles of five transitions of \ion{Si}{2} in the red part (three top panels) and in the blue part of the spectrum (two bottom panels). The solid blue line shows the absorption models, the solid green ticks show the position of the VCs, and the green shaded area shows the absorption features considered for the analysis. On top of the spectrum, residuals are shown with a (red) dashed line, and the two horizontal dashed lines show their $\pm\,1\sigma$ boundaries. The velocity scale is centred at redshift $z = 2.689959$. \label{si2}}
\end{figure}

\begin{figure}
\centering
\includegraphics[scale=0.5]{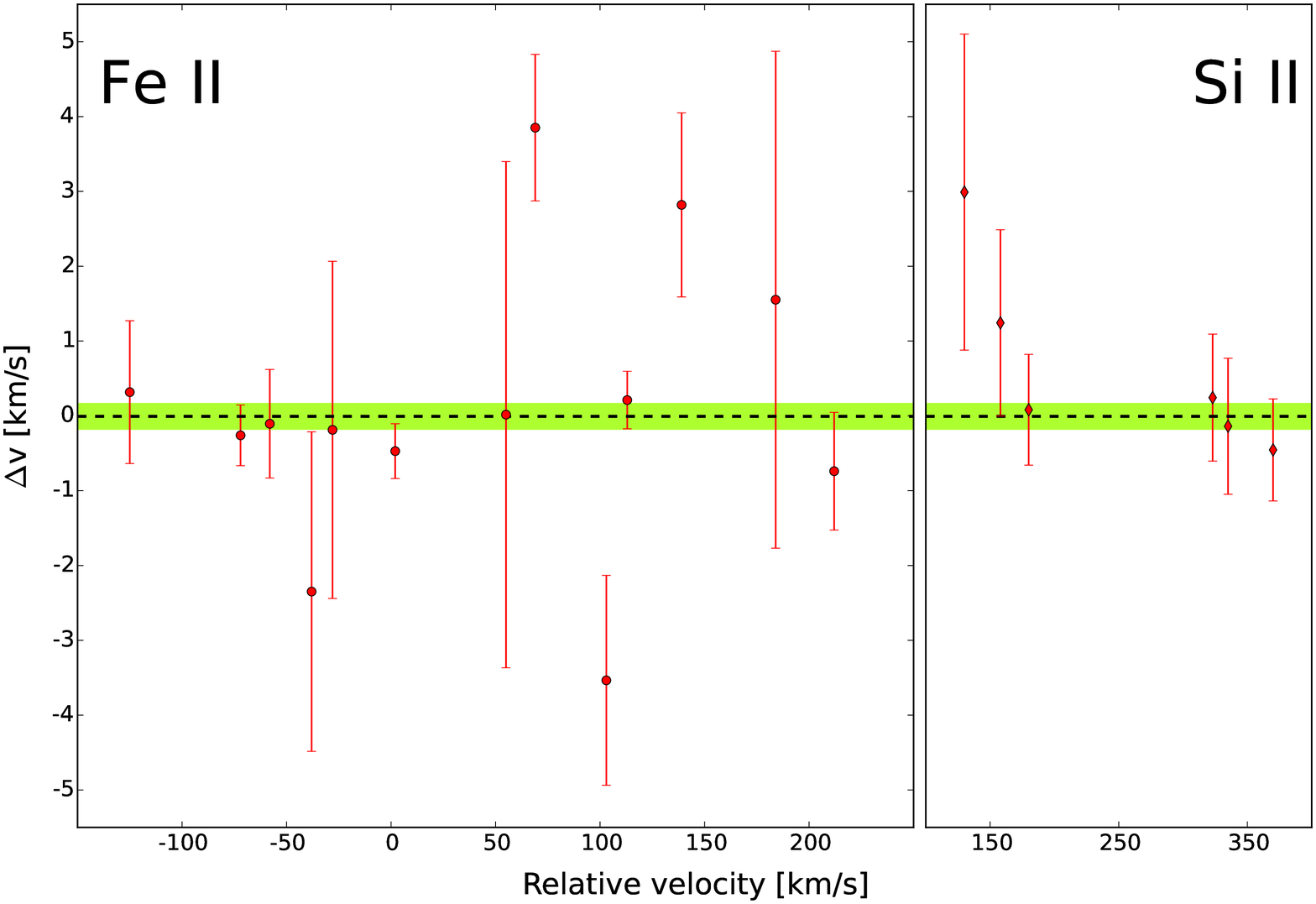}
\caption{Calculated velocity offset \mbox{$\Delta v = v_{\textrm{\scriptsize blue}}-v_{\textrm{\scriptsize red}}$} between the blue and the red arm in UVES for two atomic species. Left panel: offsets between 13 VCs of \ion{Fe}{2} at \mbox{$\zabs = 2.689570$}. Right panel: offsets between 6 VCs of \ion{Si}{2} at \mbox{$\zabs = 2.689959$}. The velocity scales are relative to these absorption redshifts. The dashed line shows the weighted average offset between the two arms and the (light green) shaded area represents its $1\sigma$ boundaries.  \label{arms_offset}}
\end{figure}

\ion{Fe}{2} is found in the absorbing system at $z \simeq 2.69$, the same that contains CO, and it has a velocity profile composed by 13 velocity components (VC) spanning \mbox{$\sim 400$ \kms}, as presented in \mbox{Fig. \ref{fe2}}. \ion{Si}{2} absorption occurs at the same redshift. Three transitions are detected in the red part of the spectrum, in the range \mbox{$\lambda = 4647 - 5644$ \AA}, and two in the blue part in the range \mbox{$\lambda = 4389 - 4409$ \AA}. In \mbox{Fig. \ref{si2}}, the broad profile of \mbox{$\sim 600$ \kms}, featuring $\sim 20$ VCs, is plotted. Almost all the \ion{Si}{2} lines, particularly those detected in the blue arm, between $-200$ and \mbox{$+130$ \kms} are heavily saturated forming a broad absorption feature that is excluded in this analysis, since it results in poorly fitted redshifts. Therefore only the 6 VCs detected between $+130$ and \mbox{$+400$ \kms} were considered for \ion{Si}{2}. The transitions considered are listed in \mbox{Table \ref{metals}}. For each element, each VC was modelled using a set of free parameters (\emph{N}, \emph{z}, \emph{b}) in \textsc{vpfit}, that were tied together among transitions detected in the same arm of UVES. This results in two outputs, one relative to the blue arm and one to the red. The fitted redshift values relative to each UVES arm were translated into relative velocities and compared exploiting any velocity shift between the two arms of UVES \mbox{$\Delta v = v_{\textrm{\scriptsize blue}} - v_{\textrm{\scriptsize red}} = [\Delta \zabs / (1+\zabs)] c$}. Here $\Delta \zabs$ is the difference between the fitted redshift value in the blue and in the red arm, and $\zabs$ is the redshift at which the absorption of the considered element occurs. The results of this comparison are presented in \mbox{Fig. \ref{arms_offset}}. \ion{Fe}{2} absorption returns an average value for the offset of \mbox{$\Delta v(\textrm{\ion{Fe}{2}}) = -0.04 \pm 0.19$ \kms} and \ion{Si}{2} returns \mbox{$\Delta v(\textrm{\ion{Si}{2}}) = 0.11 \pm 0.37$ \kms}, while the weighted average of these two offsets is \mbox{$\Delta v = -0.01 \pm 0.17$ \kms}. The main contributors to this value are the unblended \ion{Fe}{2} VCs, which have smaller errors on their \emph{z} parameters and are better described by the absorption model than the blended \ion{Fe}{2} VCs at \mbox{$\sim 60$ - 150 \kms}.
Note that the \ion{Fe}{2} and \ion{Si}{2} transitions reported by \cite{Murphy2014} have laboratory wavelength uncertainties in velocity space of \mbox{$\delta v = 14.5$} and 3.2 \ms\ respectively, while the transitions reported by \cite{Morton1991} have uncertainties \mbox{$\delta v \sim 300$ \ms}. These errors were added in quadrature to the statistical redshift errors returned from the fit. In conclusion, this analysis shows that there is no evidence of any spurious effect on $\dmm$ introduced by a combined analysis of CO absorption features detected using different arms in UVES for the J1237+0647 spectrum considered.
\begin{deluxetable}{llc}
\tablecaption{Metal transitions. \label{metals}}
\tablewidth{0pt}
\tablehead{
\colhead{Ion} & \colhead{Rest wavelength [\AA]} & \colhead{Reference} 
 }
\startdata
\ion{Fe}{2} & 1144.9379(1) & \cite{Morton1991} \\
\ion{Fe}{2} & 1608.450852(78) & \cite{Murphy2014} \\
\ion{Si}{2} & 1190.4158(1) & \cite{Morton1991}\\
\ion{Si}{2} & 1193.2897(1) & \cite{Morton1991}\\
\ion{Si}{2} & 1260.4221(1) & \cite{Morton1991} \\
\ion{Si}{2} & 1304.3702(1) & \cite{Morton1991} \\
\ion{Si}{2} & 1526.706980(16) & \cite{Murphy2014} \\
\enddata
\end{deluxetable}

\subsection{Lack of attached ThAr calibrations} \label{subsec:thar}
The lack of attached ThAr calibrations on some exposures can introduce an error on $\dmm$ of up to \mbox{$0.7 \times 10^{-6}$} \citep{Bagdonaite2014}, however no evidence for a shift of $\mu$ was found in a previous analysis of the H$_{2}$ absorption in J1237+0647 \citep{Dapra2015}. 

Only half of the dataset used in this work, that from 2009, has attached ThAr calibrations, while the other half was calibrated using the standard ThAr exposure taken at the end of the night. The impact of the lack of dedicated attached ThAr calibrations on the final value of $\dmm$ was evaluated by dividing the dataset in two subsets, one containing only the exposures taken in 2009 and the other containing the exposure taken in 2013 and 2014. From these subsets, two different constraints were retrieved: \mbox{$\dmm = (0.1 \pm 2.5_{stat}) \times 10^{-5}$} for 2009 and \mbox{$\dmm = (1.9 \pm 2.3_{stat}) \times 10^{-5}$} for 2013 and 2014. The two values agree within their uncertainties, showing that the lack of dedicated attached ThAr calibrations does not have any significant impact on the final value of $\dmm$ derived from CO absorption only presented here.

\subsection{Spectral redispersion} \label{subsec:others}
Another potential source of systematic uncertainties is spectral redispersion, caused by the co-addition of the single exposures. This implies a rebinning of the spectra on a common wavelength scale and it can distort the line-profile shapes, possibly causing a shift in $\mu$. \cite{King2011} investigated the magnitude of this effect on a similar absorbing system, and they estimated the error on $\dmm$ from 83 H$_{2}$ and HD absorption features as \mbox{$\sim 1.4 \times 10^{-6}$}. Scaling this value to the 36 CO transitions that are effectively contributing to the signal returns an error on $\dmm$ of \mbox{$\sim 2.0 \times 10^{-6}$}.

\subsection{Total systematic uncertainty} \label{subsec:total_syst}
The total systematic uncertainty on the fiducial value of the combined constraint on $\mu$ was calculated by adding all the contributions from the aforementioned sources in quadrature. This returns a systematic error of \mbox{$\sim  5.5 \times 10^{-6}$}. The fiducial constraint on $\dmm$ derived from CO absorption only therefore becomes \mbox{$\dmm = (0.7 \pm 1.6_{stat} \pm 0.5_{syst}) \times 10^{-5}$}. This value is delivered by the analysis of  CO absorption features, divided among 13 bands, from the spectrum of quasar J1237+0647 obtained by combining distortion-corrected exposures taken in 2013 and 2014 and uncorrected exposures taken in 2009.

\section{Combined analysis with H$_{2}$ and CO} \label{sec:combined}
\cite{Dapra2015} performed an independent analysis of the J1237+0647 absorption system using molecular hydrogen to constrain $\mu$ variation. Analyzing only the H$_{2}$ absorption, they found a constraint of \mbox{$\dmm = (-5.4 \pm 6.3_{stat} \pm 4.0_{syst}) \times 10^{-6}$}, which is in agreement with the constraint obtained from CO absorption only. The CO absorption is associated with the strongest velocity component of the H$_{2}$ absorption feature at \zabs=2.689551(1). The difference in redshift between this H$_{2}$ component and the CO absorption features translates into a velocity shift of \mbox{$\sim 1.2$ \kms} between the two absorbing clouds where the molecular absorption originates. Since there is no evidence for a systematic velocity offset between the arm of UVES in the spectrum of J1237+0647, the shift between H$_{2}$ and CO is not considered an artifact. Moreover, it is comparable to what was found by \cite{Noterdaeme2010} and this was considered as evidence that the absorbing system is composed of several small and dense molecular clouds.

A combined analysis using H$_{2}$, HD, and CO absorption features was performed by adding the CO fitted regions to the dataset analyzed by \cite{Dapra2015}. Only the free parameter corresponding to $\dmm$ was tied among the three molecules, while the other fitting parameters were not tied. The combined analysis delivered a constraint on the $\mu$-variation of \mbox{$\dmm = (-5.6 \pm 5.6_{stat}) \times 10^{-6}$}, whose statistical uncertainty is \mbox{$\sim 10\%$} smaller than what returned from the H$_{2}$ and HD absorption only. A weighted average of the statistical uncertainties, using the inverse of the variances as weights, delivered an error of \mbox{$5.9 \times 10^{-6}$}, which is close to what obtained from the combined analysis. Assuming that the quoted systematic uncertainties between the two measurements are not correlated, the systematic error on the combined constraint was estimated using the same procedure, which delivered an error of \mbox{$3.1 \times 10^{-6}$}. The final value of the combined constraint on a varying $\mu$ derived from H$_{2}$, HD and CO absorption is \mbox{$\dmm = (-5.6 \pm 5.6_{stat} \pm 3.1_{syst}) \times 10^{-6}$}.

\section{Conclusion} \label{sec:conclusion}
An analysis of CO absorption at high redshift in order to constrain a variation of the proton-to-electron mass ratio is performed for the first time. The absorption system at \mbox{$\zabs = 2.69$} towards quasar J1237+0647 was investigated in detail for CO and H$_{2}$ molecular absorption, while metal absorption of \ion{Fe}{2} and \ion{Si}{2} was used to analyze systematic effects. CO was found in 13 bands spread in the range \mbox{$3968 - 5702$ \AA}, falling both bluewards and redwards of the {Lyman-$\alpha$} emission peak in the quasar absorption spectrum. An absorption model describing the CO features was created starting from four molecular parameters, namely the rest wavelengths $\lambda$, the oscillator strengths $f_{\jj}$, the sensitivity coefficients \emph{K}, and the damping constants $\gamma_{\nuprime J\,'}$, which were included in an updated molecular database. The model was then fitted against the quasar spectrum using the comprehensive fitting technique, that allowed to fit simultaneously all the vibrational bands using only 4 free parameters: the total CO column density $N_{col}$, the redshift \emph{z}, the Doppler width \emph{b}, and $\dmm$.

The constraint on a varying proton-to-electron mass ratio from the CO spectrum, \mbox{$\dmm = (0.7 \pm 1.6_{stat} \pm 0.5_{syst}) \times 10^{-5}$}, is less stringent than that obtained from the H$_{2}$ spectrum in this absorption system. This may in part be due to the overlap of the \bx, \cx, and \ex\ bands by broad \ion{H}{1} features, which may be absent in other absorption systems than that towards J1237+0647 analyzed here. The present paper lists the database covering the relevant molecular properties of CO electronic absorption lines that may be used in future studies searching for $\mu$-variation based on optical absorption of carbon monoxide in the line-of-sight of high redshift quasars. Thus far CO electronic absorption has been detected in six different systems towards quasars \citep{Srianand2008,Noterdaeme2008,Noterdaeme2009,Noterdaeme2010,Noterdaeme2011}, as well as in some Gamma-Ray-Burst observations \citep{Prochaska2009}, where the same methods can potentially be applied to detect $\mu$-variation.

The CO absorption was included in a combined analysis involving the previously investigated H$_{2}$ \citep{Dapra2015}, leading to a more stringent constraint on a varying $\mu$ of \mbox{$\dmm = (-5.6 \pm 5.6_{stat} \pm 3.1_{syst}) \times 10^{-6}$}. Various potential sources of systematics were investigated, including the long-range distortions that are known to affect the UVES spectra. The J1237+0647 spectrum was partially corrected for such distortions using the supercalibration technique presented by \cite{Whitmore2015}. This result is the first independent constraint on a varying $\mu$ obtained from the analysis of optical absorption for two different molecules detected in the same absorbing system, thus observed under the same physical conditions. The constraint agrees with previous results derived from 10 different systems in the range \mbox{$\zabs = 2.05 - 4.22$}, which correspond to a time interval of \mbox{$10 -12.5$ Gyrs}. They return an averaged constraint that shows a null $\mu$-variation at a level \mbox{of $\sim 5 \times 10^{-6}$} (3-$\sigma$) \citep{Ubachs2016}. 

\acknowledgments
The authors thank the Netherlands Foundation for Fundamental Research of Matter (FOM) for financial support. The work is based on observations with the ESO Very Large Telescope at Paranal (Chile). WU acknowledges funding from the European Research Council (ERC) under the European Union's Horizon 2020 research and innovation programme (grant agreement No 670168). MTM thanks the Australian Research Council for \textsl{Discovery Project} grant DP110100866 which supported this work.

{\it Facilities:} \facility{VLT (UVES)}.


\end{document}